\documentclass[12pt]{iopart}

\usepackage{graphicx,color,subfigure}
\usepackage{ae,aecompl}
\begin{document}

\title[One- and two-dimensional reductions of the mean-field description of
degenerate Fermi gases]{One- and two-dimensional reductions of the mean-field description of
degenerate Fermi gases}

\author{Pablo~D\'{i}az}
\address{Departamento de Ciencias F\'{i}sicas, Universidad de la
Frontera, Casilla 54-D, Temuco, Chile.}
\address{Departamento de F\'{i}sica, Universidad T\'{e}cnica
Federico Santa Mar\'{i}a, Casilla 110-V, Valpara\'{i}so, Chile.}
\author{David~Laroze}
\address{Max Planck Institute for Polymer Research, D-55021
Mainz, Germany}
\address{Instituto de Alta Investigaci\'{o}n, Universidad de
Tarapac\'{a}, Casilla 7D, Arica, Chile}
\ead{laroze@mpip-mainz.mpg.de}
\author{Iv\'{a}n~Schmidt}
\address{Departamento de F\'{i}sica, Universidad T\'{e}cnica
Federico Santa Mar\'{i}a, Casilla 110-V, Valpara\'{i}so, Chile}
\author{Boris A.Malomed}
\address{Department of Physical Electronics, School of
Electrical Engineering, Faculty of Engineering, Tel Aviv University,
Tel Aviv IL-69978, Israel}

\begin{abstract}
We study collective behavior of Fermi gases trapped in various external
potentials, including optical lattices (OLs), in the framework of the
mean-field (hydrodynamic) description. Using the variational method, we
derive effective dynamical equations for the one- and two-dimensional (1D
and 2D) settings from the general 3D mean-field equation. The respective
confinement is provided by trapping potentials with the cylindrical and
planar symmetry, respectively. The resulting equations are nonpolynomial Schr%
\"{o}dinger equations (NPSEs) coupled to equations for the local transverse
size of the trapped states. Numerical simulations demonstrate close
agreement of results produced by the underlying 3D equation and the
effective low-dimensional ones. We consider the ground state in these
settings. In particular, analytical solutions are obtained for the
effectively 2D non-interacting Fermi gas. Differences between the 1D and 2D
configurations are highlighted. Finally, we analyze the dependence of the 1D
and 2D density patterns of the trapped gas, in the presence of the OL, on
the strengths of the confining and OL potentials, and on the scattering
length which determines the strength of interactions between non-identical
fermions.
\end{abstract}


\section{Introduction}
\label{intro}

In recent years, ultracold atomic gases have been explored as the means of
experimental and theoretical simulations of various aspects of many-body
physics \cite{RepProgPhys}-\cite{Kartashov2011}. In particular, conventional
solid-state physics can be emulated by gases of fermionic atoms. The
possibility to adjust the depth of the optical lattices (OLs), into which
the gas is usually loaded, and the strength of interactions between atoms
via the Feshbach resonance, opens the way to a detailed experimental study
of these settings \cite{Snoek2011a}-\cite{Frohlich2011}. Such possibilities
have inspired, in particular, many works dealing with the BCS-BEC crossover
\cite{Tempere2006}-\cite{Watanabe2011}, which occurs on the route from weak
to strong attraction, being particularly relevant as a simulator of
superconductivity.

In these contexts, the density-functional theory (mean field) and the
local-density approximation have been used with great success for the
description of quantum gases close to their ground states \cite{Snoek2011b}-%
\cite{Jackson1998}, resulting in effective time-dependent nonlinear Schr\"{o}%
dinger equations. In the framework of this approach, hydrodynamic equations
for the superfluid Fermi gas at zero temperature were derived in Ref. \cite%
{Kim2004}, and, in a similar way, the hydrodynamic equations which remain
valid in the presence of temporal variations of control parameters were
derived in Ref. \cite{Capuzzi2008}.

The possibility of highly confined systems in one and two dimensions has
stimulated the search for dimensionally reduced descriptions. In particular,
quasi-one-dimensional Fermi systems described by many-body Hamiltonians have
been analyzed, following seminal works \cite{Lieb1963,Gaudin1967} and a more
recent one \cite{Fuchs2004}. A variational approach for 1D and 2D Fermi
gases in the presence of OL potentials was developed in Ref. \cite%
{Malomed2009}, and a similar technique for 1D and 2D Fermi-Fermi mixtures
was proposed still earlier in works \cite{Adhikari2006} and \cite%
{Adhikari2007b}.

In this paper we address Fermi gases in the BCS regime in nearly 2D and 1D
configurations, i.e., disk- and cigar-shaped ones, respectively. The
configurations may also include inner OL potentials. To reduce the 3D
nonpolynomial Schr\"{o}dinger equation (NPSE) for the Fermi gas, known from
the hydrodynamic description, to the corresponding 2D and 1D equations, we
use the variational method, similar to that employed in Refs. \cite%
{Salasnich2002,Salasnich2D} to reduce the 3D Gross-Pitaevskii equation for
the Bose-Einstein condensate to the respective 1D and 2D bosonic NPSEs.
Unlike variational methods usually employed in the description of confined
fermion systems, which postulate a constant width of the wave-function
profile in the confined dimensions, the present method accounts for
variations of the confinement width, resulting in significant corrections to
the density profile. We demonstrate that our improved approximation is
essentially closer to results of the 3D integration.

As a starting point, we derive an effective hydrodynamics equation for a
multicomponent Fermi gas with balanced populations, generalizing those found
in the literature for a gas with two balanced populations, considering the
BCS limit the density of energy. Next, we derive a effective equations for
the disk-shaped trap, in the form of the respective equation for the
fermionic mean-field wave function coupled to an equation for the transverse
width of the trapped state. Analytical solutions of the 2D equations can be
found for the non-interacting gas; in the presence of the interactions
between fermions with spins up and down, we develop numerical solutions, and
demonstrate that the reduced equations reproduce the results obtained from
the full 3D equation with a very good accuracy. We further extend the
analysis to include the 2D (in-plane) lattice and superlattice potentials.
Then, we examine the reduction of the 3D NPSE to 1D equations in the case of
the cigar-shaped confinement. The quasi-1D configuration including the axial
OL potential is investigated too.

\section{The model}

\label{sec:1}

\subsection{Effective hydrodynamic equation for two balanced populations}

We consider a weakly interacting dilute Fermi gas of $N$ atoms of mass $m$
and atomic spin $s$, which form the BCS superfluid, loaded into an optical
trap. The gas is taken at zero temperature, with balanced populations of the
$2s+1$ possible spin states, as specified in detail below. To do this, we
start with the case of balanced populations, for which an effective
hydrodynamic equation was derived from the time-dependent density-functional
theory for the Fermi gas, in the regime of the BCS-BEC crossover, by Kim and
Zubarev in Ref. \cite{Kim2004}:

\begin{equation}
i\hbar {\partial _{t}}\Psi \left( {\mathbf{r},t}\right) =\left[ {\ -\frac{{%
\hbar ^{2}}}{{2m}}{\nabla ^{2}}+V\left( {\mathbf{r}}\right) +\mu \left(
n\left( {\mathbf{r},t}\right) \right) }\right] \Psi \left( {\mathbf{r},t}%
\right) ,  \label{E-b0}
\end{equation}%
where $\Psi $ is the superfluid wave function, such that $n\left( {\mathbf{r}%
,t}\right) ={\left\vert \Psi \left( {\mathbf{r},t}\right) \right\vert ^{2}}$
is the particle density, and
\begin{equation}
\mu \left( n\right) =\frac{\partial }{\partial n}\left[
{n\varepsilon \left( n\right) }\right]   \label{mu}
\end{equation}%
is the chemical potential, related to the energy per particle
$\varepsilon \left( n\right) $. In Ref. \cite{Kim2004} it was also
shown, for elongated cigar-shaped potential traps, that the von
Weizs\"{a}cker's correction to the Thomas-Fermi kinetic-energy
density (which corresponds to the hydrodynamic approximation)
\cite{Weizsacker1935}, ${\varepsilon _{\mathrm{TF}}}\left( n\right)
=\left( {3/10}\right) n{\hbar ^{2}}{\left( {3{\pi ^{2}}n}\right)
^{2/3}}/m$,
may be important even for the gas with a large density of fermions, $%
\left\vert {\nabla n}\right\vert /{n^{4/3}}\ll 1$ (this is the applicability
condition for the hydrodynamic approximation). Such an improved hydrodynamic
approximation corresponds to the first two terms of the Kirzhnitz's gradient
expansion \cite{Kirzhnitz1967}. For large smoothly varying densities in the
free space, the respective correction is negligible; however, for strongly
confined systems the additional term (the kinetic energy) may be comparable
to the main one, hence it cannot be neglected, as shown below by our
variational approach.

A problem also arises at edges of the trapped superfluid, where $\left\vert {%
\nabla n}\right\vert /{n^{4/3}}$ ceases to be small. To overcome
this problem, one may consider the trapped gas with large number of
particles, lessening the relative significance of the edges. The
case of a relatively small number of particles ($N\sim 10^{2}$) is
considered below too, just with the aim to test the accuracy of the
variational method.

\subsection{The BCS regime}

For a fermionic gas composed of two spin states with balanced populations
and a negative scattering length of the $s$-wave collisions between the
particles in different spin states, ${a_{s}}<0$, the BCS limit corresponds
to $k_{F}{\left\vert {a_{s}}\right\vert }\ll 1$, with the Fermi wavenumber $%
k_{F}=(3\pi ^{2}n)^{1/3}$. In this regime, the energy per particle, $%
\varepsilon $, is represented by the expansion in powers of $k_{F}{\left\vert {%
a_{s}}\right\vert }$ \cite{Huang1957}:

\begin{equation}
\varepsilon \left( n\right) =\frac{3}{5}{\varepsilon _{F}}\left[ {1+\frac{{10}}{{%
9\pi }}{k_{F}}{a_{s}}+\frac{{4\left( {11-2\ln \left( 2\right) }\right) }}{{21%
{\pi ^{2}}}}{{\left( {{k_{F}}{a_{s}}}\right) }^{2}}+\cdots }\right] ,
\label{E-b1}
\end{equation}%
where ${\varepsilon _{F}}={\hbar ^{2}}k_{F}^{2}/\left( {2m}\right) $
is the Fermi energy. From Eqs. (\ref{E-b1}) and (\ref{mu}) it
follows that the chemical potential takes the form

\begin{equation}
\mu \left( n\right) =\frac{{\hbar ^{2}}}{{2m}}{\left( {3{\pi ^{2}}}\right)
^{2/3}}{n^{2/3}}+\frac{{2{\hbar ^{2}}\pi {a_{s}}}}{m}n\left[ {1+{1.893a_{s}}{%
n^{1/3}}+\cdots }\right]   \label{E-b2}
\end{equation}%
where the first term corresponds to the effective Pauli repulsion, and the
following ones to the superfluidity due to collisions between the fermions
in different spin states. Substituting the latter expression in Eq.(\ref%
{E-b0}), and keeping only the first collisional term, we obtain the known
nonlinear Schr\"{o}dinger equation for the fermionic superfluid (see, e.g.,
Refs. \cite{Kim2004,Adhikari}),
\begin{equation}
i\hbar {\partial _{t}}\Psi =\left[ {\ -\frac{{\hbar ^{2}}}{{2m}}{\nabla ^{2}}%
+V+\frac{{\hbar ^{2}}}{{2m}}{{\left( {3{\pi ^{2}}}\right) }^{2/3}}{n^{2/3}}+%
\frac{{2\pi {\hbar ^{2}}{a_{s}}}}{m}n}\right] \Psi ,  \label{E-b3}
\end{equation}%
where the last term is similar to that in the Gross-Pitaevskii equation for
bosons, but with an extra factor of $1/2$, as the Pauli exclusion principle
allows only atoms in different spin states interact via the scattering.
Equation (\ref{E-b3}) implies equal particle densities and phases of the
wave functions associated with both spin states.

\subsection{The effective hydrodynamic equation for multiple balanced
populations}

We now aim to extrapolate the previous case to systems with multiple atomic
spin states, $\sigma _{j}$, associated with $2s+1$ values of the vertical
projection of the spin. To this end, we treat the atoms in each spin state ($%
\sigma _{j}$) as a fully polarized Fermi gas, and include interactions
between atoms in the different spin state (assuming a single value of the
scattering length, $a_{s}$), which leads to the following
system of equations:
\begin{eqnarray}
i\hbar {\partial _{t}}{\Psi _{j}}\left( {\mathbf{r},t}\right)  &=&\left[ {\ -%
\frac{{\hbar ^{2}}}{{2m}}{\nabla ^{2}}+V\left( {\mathbf{r}}\right) +\frac{{%
\hbar ^{2}}}{{2m}}{{\left( {6{\pi ^{2}}}\right) }^{2/3}}{{n_{j}\left( {%
\mathbf{r},t}\right) }^{2/3}}}\right] {\Psi _{j}}\left( {\mathbf{r},t}%
\right)   \nonumber \\
&&+\frac{{4\pi {\hbar ^{2}}{a_{s}}}}{m}\sum\limits_{k\neq j=-(s+1/2)}^{s+1/2}{{{n_{k}\left( {%
\mathbf{r},t}\right) }}{\Psi _{j}}}\left( {\mathbf{r},t}\right) ,
\label{E-1}
\end{eqnarray}%
where $\Psi _{j}$ is the wave function of the superfluid associated with
spin projection $\sigma _{j}$, such that $n_{j}\left( {\mathbf{r},t}\right) =%
{\left\vert {{\Psi _{j}}\left( {\mathbf{r},t}\right) }\right\vert ^{2}}$ is
the respective particle density, and $V(\mathbf{r})$ an external potential,
which is assumed to be identical for all the spin states. A relevant example
of a higher nuclear spin, $s=9/2$ (decoupled from the electron states), in a
degenerate Fermi gas of $^{87}$Sr, was demonstrated in Ref. \cite{Tey2010}.

In the case of fully locally balanced populations, the density of particles
is the same in each component, $n_{1}=n_{2}=...=n_{2s+1}$, hence the total
density is $n=(2s+1)n_{j}$. Assuming also equal phases of the wave-function
components, we define a single wave function, $\Psi =\sqrt{2s+1}\Psi _{j}$,
hence Eq. (\ref{E-1}) is replaced by

\begin{equation}
i\hbar {\partial _{t}}{\Psi }=\left[ -\frac{{{\hbar ^{2}}}}{{2{m}}}{\nabla
^{2}}+{V}+\frac{{\hbar ^{2}}}{{2m}}{{\left( {\frac{{6{\pi ^{2}}}}{{2s+1}}}%
\right) }^{2/3}}n_{\mathrm{3D}}^{2/3}+g{n_{\mathrm{3D}}}\right] {\Psi },
\label{E-2}
\end{equation}%
with density ${n_{\mathrm{3D}}}\left( {{\mathbf{r}},t}\right) ={\left\vert {%
\Psi \left( {{\mathbf{r}},t}\right) }\right\vert ^{2}}$ and scattering
coefficient%
\begin{equation}
g\equiv 8s\pi {\hbar ^{2}}{a_{s}}/(2s+1)m.  \label{E-3}
\end{equation}%
In particular, the fully polarized gas, with the interactions between
identical fermions suppressed by the Pauli principle, formally corresponds
to $s=0$ (hence, $g=0$, as given by Eq. (\ref{E-3})). Unless specified
otherwise, we set $s=1/2$ below, focusing on the case of balanced
populations in two components.

\subsection{The Lagrangian density}

Equation (\ref{E-2}) can be derived, as the Euler-Lagrange equation,

\begin{equation}
\frac{{\delta \mathcal{L}}}{{\delta {\Psi ^{\ast }}}}=\frac{{\partial
\mathcal{L}}}{{\partial {\Psi ^{\ast }}}}-\frac{\partial }{{\partial t}}%
\frac{{\partial \mathcal{L}}}{{\partial \left( {{\partial _{t}}{\Psi ^{\ast }%
}}\right) }}-\nabla \frac{{\partial \mathcal{L}}}{{\partial \left( {\nabla {%
\Psi ^{\ast }}}\right) }}=0,  \label{E-5}
\end{equation}%
from the corresponding action, $\mathcal{S}=\int {dtd{\mathbf{r}}\mathcal{L}}
$, with the Lagrangian density
\begin{eqnarray}
\mathcal{L} &=&i\frac{\hbar }{2}\left( {{\Psi ^{\ast }}\frac{{\partial \Psi }%
}{{\partial t}}-\Psi \frac{{\partial {\Psi ^{\ast }}}}{{\partial t}}}\right)
-\frac{{{\hbar ^{2}}}}{{2{m}}}{\left\vert {\nabla {\Psi }}\right\vert ^{2}}-{%
V}(\mathbf{r}){n_{\mathrm{3D}}}-  \nonumber \\
&&\frac{{{\hbar ^{2}}}}{{2{m}}}\frac{3}{5}{\left( {\frac{{6{\pi ^{2}}}}{{2{s}%
+1}}}\right) ^{2/3}}n_{\mathrm{3D}}^{5/3}-\frac{1}{2}{g}n_{\mathrm{3D}}^{2},
\label{E-4}
\end{eqnarray}%
where the asterisk stands for the complex conjugate. Similar Lagrangian
formalisms have been used, in the context of the density-functional theory,
in diverse settings \cite{Adhikari,Adhikari2007b,Kim2004a}.

Our goal is to reduce the 3D description to one and two dimensions, using
the variational approach. As said above, this derivation may be compared to
that for the low-dimensional NPSE from the 3D Gross-Pitaevskii equation for
the bosonic gas \cite{Salasnich2002,Salasnich2D}. The approach is based on
the factorization of the 3D wave function, substituting the corresponding
ansatz into the action, and integrating out the factor corresponding to the
direction(s) orthogonal to the confined dimensions.

\section{The two-dimensional reduction}
\label{sec:2}
\subsection{The derivation of the 2D equations}

Here we aim to derive effective 2D equations for the Fermi gas in a
disk-shaped trap. For this purpose, we consider the external potential
composed of two terms: the parabolic (harmonic-oscillator) one accounting
for the confinement in the $z$ direction, transverse to the disk's plane,
and the in-plane potential, $V_{\mathrm{2D}}$:

\begin{equation}
V\left( {\mathbf{r}}\right) =\frac{1}{2}m\omega _{z}^{2}{z^{2}+}V_{\mathrm{2D%
}}\left( {x,y}\right) .  \label{E-7}
\end{equation}%
As said above, the initial ansatz assumes the factorization of the 3D wave
function into a product of functions of $z$ and $\left( x,y\right) $, the
former one being the ground state of the harmonic-oscillator potential \cite%
{Jackson1998}:

\begin{equation}
\Psi \left( {{\mathbf{r}},t}\right) =\frac{1}{{{\pi ^{1/4}}}\sqrt{{\xi {{%
\left( {x,y,t}\right) }}}}}{\exp }\left( -\frac{{{z^{2}}}}{{2}\left( {\xi {{%
\left( {x,y,t}\right) }}}\right) ^{2}}\right) \phi \left( {x,y,t}\right) ,
\label{E-8}
\end{equation}%
subject to the unitary normalization, with transverse width $\xi (z,t)$
considered as a variational function, while the 2D wave function, $\phi $,
is normalized to the number of atoms. Therefore, the reduction from 3D to 2D
implies that the system of equations should be derived for the pair of
functions $\phi \left( x,y,t\right) $ and $\xi \left( x,y,t\right) $, using
the reduced action functional to be derived by integrating the 3D action
over the $z$-coordinate (cf. the derivation for the bosonic gas \cite%
{Salasnich2002,Salasnich2D}):

\begin{equation}
\mathcal{S}{_{\mathrm{2D}}}=\int {dtdxdy{\mathcal{L}}_{\mathrm{2D}}},
\label{E-9}
\end{equation}%
where the respective Lagrangian density is

\begin{eqnarray}
{\mathcal{L}_{\mathrm{2D}}} &=&i\frac{\hbar }{2}\left( {{\phi ^{\ast }}{%
\partial _{t}}\phi -\phi {\partial _{t}}{\phi ^{\ast }}}\right) -\frac{{%
\hbar ^{2}}}{{2m}}{\left\vert {{\nabla _{\bot }}\phi }\right\vert ^{2}}-{V_{%
\mathrm{2D}}n_{\mathrm{2D}}}  \nonumber \\
&&-\left[ \frac{{\hbar ^{2}}}{{2m}}\frac{{3C_{\mathrm{2D}}}}{{5{\xi ^{2/3}}}}%
n_{{\mathrm{2D}}}^{5/3}+\frac{g}{{2{{\left( {2\pi }\right) }^{1/2}}\xi }}n_{{%
\mathrm{2D}}}^{2}+\frac{{\hbar ^{2}}}{{4m{\xi ^{2}}}}{n_{{\mathrm{2D}}}}+%
\frac{{m\omega _{z}^{2}{\xi ^{2}}}}{4}{n_{{\mathrm{2D}}}}\right] ,
\label{E-10}
\end{eqnarray}%
with 2D density ${n_{\mathrm{2D}}}\left( x,y\right) ={\left\vert \phi \left(
x,y\right) \right\vert ^{2}}$, $C_{\mathrm{2D}}\equiv {%
(3/5)^{1/2}(6/(2s+1))^{2/3}}\pi $. The last two terms in Eq. (\ref{E-10})
result from the reduction to 2D: the first among them shows that stronger
confinement implies a higher energetic cost, while the quadratic dependence
on $\xi $ in the last term originates from the original confining potential
acting in the $z$ direction. These two terms are relevant for configurations
with an inhomogeneous spatial density, while for spatially homogeneous
solutions they may be omitted.

\begin{table}[tbh]
\begin{center}
\begin{tabular}{|c|c|c|c|}
\hline
Atoms & $\mathop {\omega _{c} \mathrm{[Hz]} }\limits_{a_{c}=0.1\times {%
10^{-6}}\mathrm{m}}$ & $\mathop {\omega _{c} \mathrm{[Hz]}}\limits_{a_{c}=1.0%
\times {10^{-6}}\mathrm{m}}$ & $\mathop {\omega _{c} \mathrm{[Hz]}}%
\limits_{a_{c}=10.0\times {10^{-6}}\mathrm{m}}$ \\ \hline
$^{6}$Li & $10.5\times 10^{5}$ & $10.5\times 10^{3}$ & $105.$ \\
$^{40}$K & $1.58\times 10^{5}$ & $1.58\times 10^{3}$ & $15.8$ \\
$^{84}$Sr & $0.75\times 10^{5}$ & $0.75\times 10^{3}$ & $7.5$ \\ \hline
\end{tabular}%
\end{center}
\caption{Relevant values of the characteristic frequency, $\protect\omega %
_{c}$, for $^{6}$Li, $^{40}$K and $^{84}$Sr, assuming three possible values
of the characteristic length, $a_{c}$.}
\label{Table-1}
\end{table}

To present the effective 2D equations in a dimensionless form, we define a
characteristic length, $a_{c}$, and a characteristic frequency, $\omega _{c}$%
, related through the atomic mass ($m$), $\omega _{c}=\hbar /(ma_{c}^{2})$.
In this way, the physical units of the system depend on fix the value of $%
a_{c}$ (or $\omega _{c}$) for particular atomic species, see Table \ref%
{Table-1}. Thus, we rescale the variables and constants as $\tau \equiv
\omega _{c}t/2$, $\left\{ X,Y\right\} \equiv \left\{ x,y\right\} /a_{c}$, $%
\vartheta \equiv {a_{c}}\phi $, ${\rho _{\mathrm{2D}}}\equiv {\left\vert
\vartheta \right\vert ^{2}}$, $\zeta \equiv \xi /{a_{c}}$, $U_{\mathrm{2D}%
}\equiv 2V_{\mathrm{2D}}/(\hbar \omega _{c})$, $G_{\mathrm{2D}}\equiv {%
2^{7/2}}\sqrt{\pi }s{a_{s}}/\left[ {(2s+1)a_{c}}\right] $, and $\varpi
_{z}\equiv \omega _{z}/\omega _{c}$. In this notation, time $t$ is expressed
in units of $2/\omega _{c}$, and 2D coordinates $\left\{ {x,y}\right\} $ in
units of $a_{c}$. Then, Eqs. (\ref{E-9}) and (\ref{E-10}) are transformed
into

\begin{equation}
\mathcal{S}=\hbar \int {d\tau dXdY{\tilde{\mathcal{L}}}_{\mathrm{2D}}}
\label{E-11}
\end{equation}%
and
\begin{eqnarray}
{{{\tilde{\mathcal{L}}}}_{\mathrm{2D}}} &=&\frac{i}{2}\left( {\vartheta
^{\ast }{\partial _{\tau }}\vartheta -\vartheta {\partial _{\tau }}\vartheta
^{\ast }}\right) -{\left\vert {\nabla \vartheta }\right\vert ^{2}}-{U_{%
\mathrm{2D}}\rho _{\mathrm{2D}}}  \nonumber \\
&&-\left[ \frac{3}{{5{\zeta ^{2/3}}}}C_{\mathrm{2D}}\rho _{\mathrm{2D}%
}^{5/3}+\frac{G_{\mathrm{2D}}}{{2\zeta }}\rho _{2D}^{2}+\frac{1}{{2{\zeta
^{2}}}}{\rho _{\mathrm{2D}}}+\frac{1}{2}\varpi _{z}^{2}{\zeta ^{2}}{\rho _{%
\mathrm{2D}}}\right] ,  \label{E-12}
\end{eqnarray}%
where the part in the square brackets corresponds to the local
energy density, $\varepsilon _{\mathrm{2D}}$, when the system is in
a spatially homogeneous state in the absence of external potentials:

\begin{equation}
{{\varepsilon }_{\mathrm{2D}}}=\frac{3}{{5{\zeta ^{2/3}}}}C_{\mathrm{2D}}\rho _{%
\mathrm{2D}}^{5/3}+\frac{G_{\mathrm{2D}}}{{2\zeta }}\rho _{\mathrm{2D}}^{2}+%
\frac{1}{{2{\zeta ^{2}}}}{\rho _{\mathrm{2D}}}+\frac{1}{2}\varpi _{z}^{2}{%
\zeta ^{2}}{\rho _{\mathrm{2D}}}~.  \label{E-13}
\end{equation}%
The latter equation is a relation between the energy and particle densities,
demonstrating how different interaction terms affect the local stability of
the matter wave.

The Euler-Lagrange equations produced by varying action $\mathcal{S}$ with
respect to $\vartheta $ and $\zeta $ take the form of

\begin{equation}
i{\partial _{\tau }}\vartheta =\left[ -\nabla _{\bot }^{2}+{U_{\mathrm{2D}}}+%
{\frac{1}{{{\zeta ^{2/3}}}}C_{\mathrm{2D}}{{\left\vert \vartheta \right\vert
}^{4/3}}+\frac{G_{\mathrm{2D}}}{\zeta }{{\left\vert \vartheta \right\vert }%
^{2}}+\frac{1}{{2{\zeta ^{2}}}}+\frac{{\varpi _{z}^{2}}}{2}{\zeta ^{2}}}%
\right] \vartheta ,  \label{E-14}
\end{equation}

\begin{equation}
\varpi _{z}^{2}{\zeta ^{4}}-\frac{2}{5}{C_{\mathrm{2D}}\left\vert \vartheta
\right\vert ^{4/3}}{\zeta ^{4/3}}-\frac{1}{2}{G_{\mathrm{2D}}\left\vert
\vartheta \right\vert ^{2}}\zeta -1=0,  \label{E-15}
\end{equation}%
cf. Eq.~(\ref{E-5}). Algebraic equation (\ref{E-15}) for $\zeta $ cannot be
solved analytically, therefore we used the Newton method to solve it
numerically. The necessity to find $\zeta $ at each step of the integration
is a numerical complication of a minimal cost compared to the 3D integration
of the underlying equation (\ref{E-2}).

Nevertheless, for the non-interacting gas, with $G_{\mathrm{2D}}=0$, Eq.~(%
\ref{E-15}) is a cubic equation for $\zeta ^{4/3}$, which admits an
analytical solution, the single physically relevant real root being
\begin{equation}
{\varpi _{z}^{2/3}}{\zeta ^{4/3}}\left( \Omega \right) =\frac{{2}^{1/3}{%
\Omega }}{{3}\left( {\sqrt{1-\frac{4}{{27}}{\Omega ^{3}}}+1}\right) ^{1/3}}+%
\frac{\left( {\sqrt{1-\frac{4}{{27}}{\Omega ^{3}}}+1}\right) ^{1/3}}{{2}%
^{1/3}},  \label{E-16}
\end{equation}%
where $\Omega \equiv 2C_{\mathrm{2D}}\left\vert \vartheta \right\vert
^{4/3}/(5\varpi _{z}^{2/3})$. This root is real at ${\rho _{\mathrm{2D}}}/{%
\varpi _{z}}<{5^{9/4}}(2s+1)/({2^{7/2}}{3^{1/4}}{\pi ^{3/2}})\approx
0.45(2s+1)$, i.e., for sufficiently low densities and the strong transverse
confinement. Further, in the low-density limit, the approximation for the
root is
\begin{equation}
\zeta \approx \frac{1}{{{\varpi _{z}^{1/2}}}}\left( {1+\frac{C_{\mathrm{2D}}%
}{{10\varpi _{z}^{2/3}}}{{\left\vert \vartheta \right\vert }^{4/3}}}\right) ,
\label{E-17}
\end{equation}%
the substitution of which into Eq.~(\ref{E-14}) results in the following
NPSE for $\vartheta $:

\begin{equation}
i{\partial _{\tau }}\vartheta =\left[ {-\nabla _{\bot }^{2}+{U}_{\mathrm{2D}%
}+\varpi _{z}^{1/3}C_{\mathrm{2D}}\left( {{{\left\vert \vartheta \right\vert
}^{4/3}}-\frac{{C_{\mathrm{2D}}{{\left\vert \vartheta \right\vert }^{8/3}}}}{%
{15\varpi _{z}^{2/3}}}}\right) }\right] \vartheta ,  \label{E-18}
\end{equation}%
where the chemical potential is shifted by a constant term, $2\varpi _{z}$.
In the limit of ${\left\vert \vartheta \right\vert \rightarrow 0}$, Eq. (\ref%
{E-17}) yields $\zeta ={{\varpi _{z}^{-1/2}}}$, which corresponds to the
width of the ground state of the harmonic oscillator.

\begin{figure}[tbp]
\centering
\resizebox{\textwidth}{!}{
\includegraphics{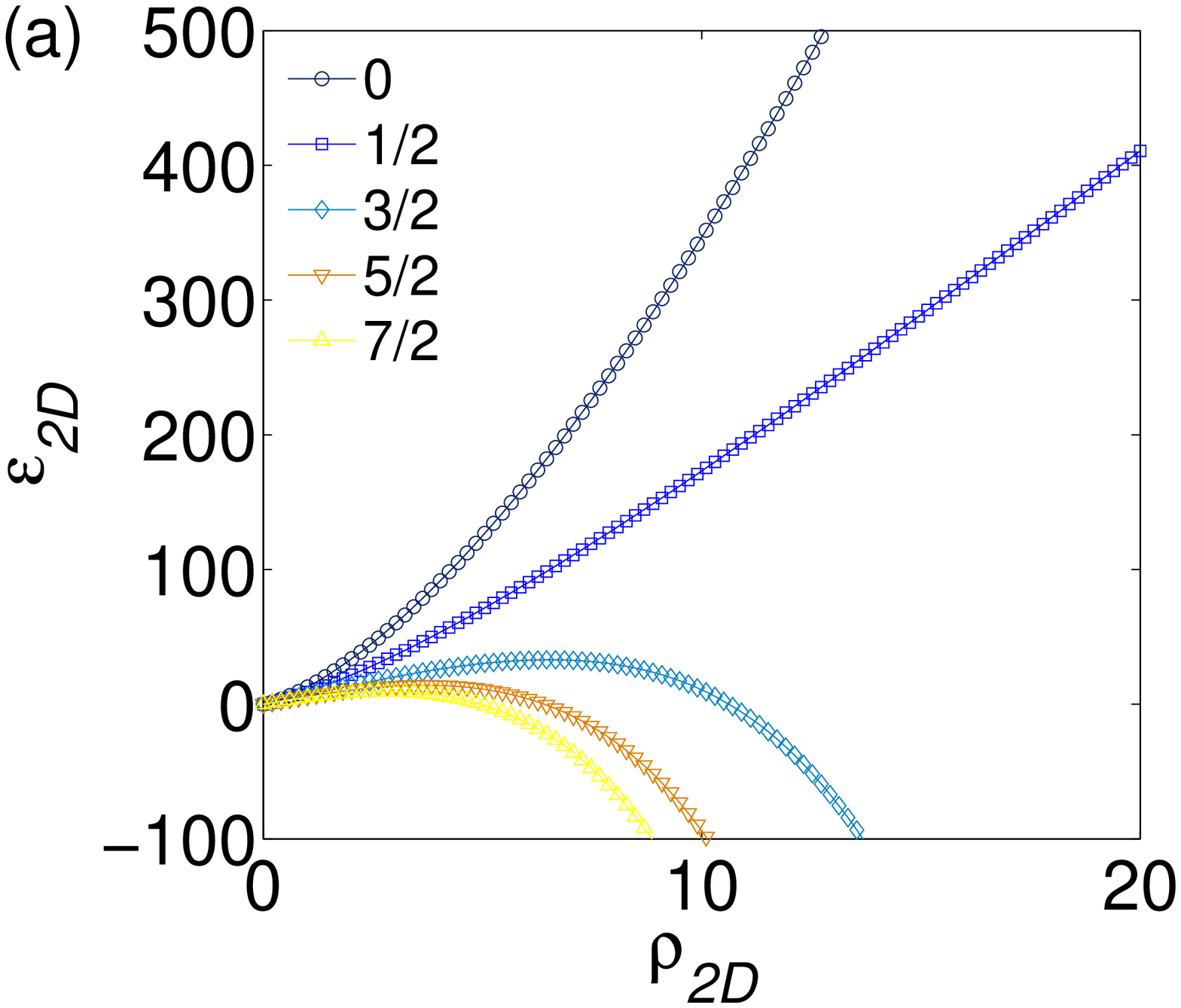}
\includegraphics{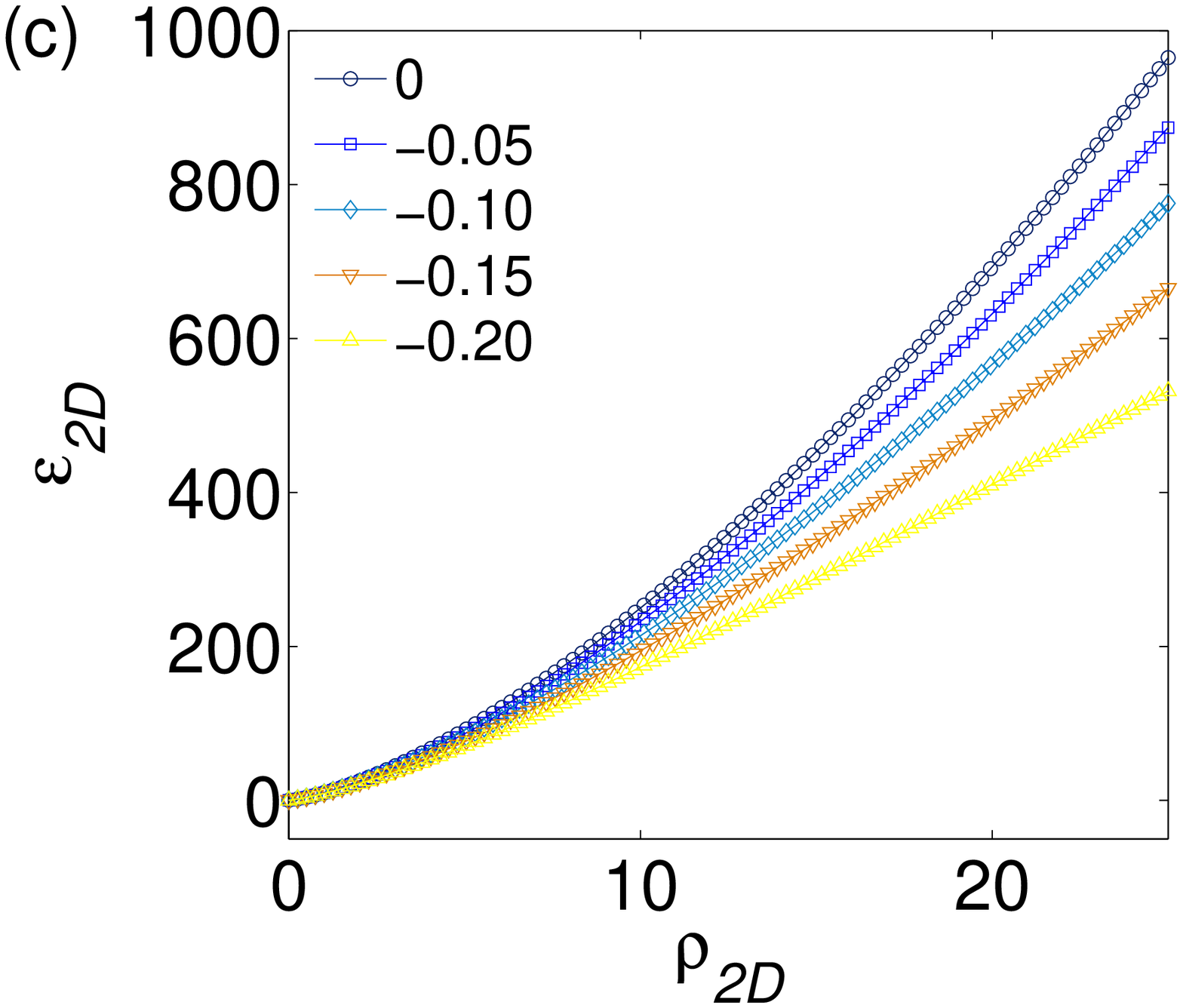}
\includegraphics{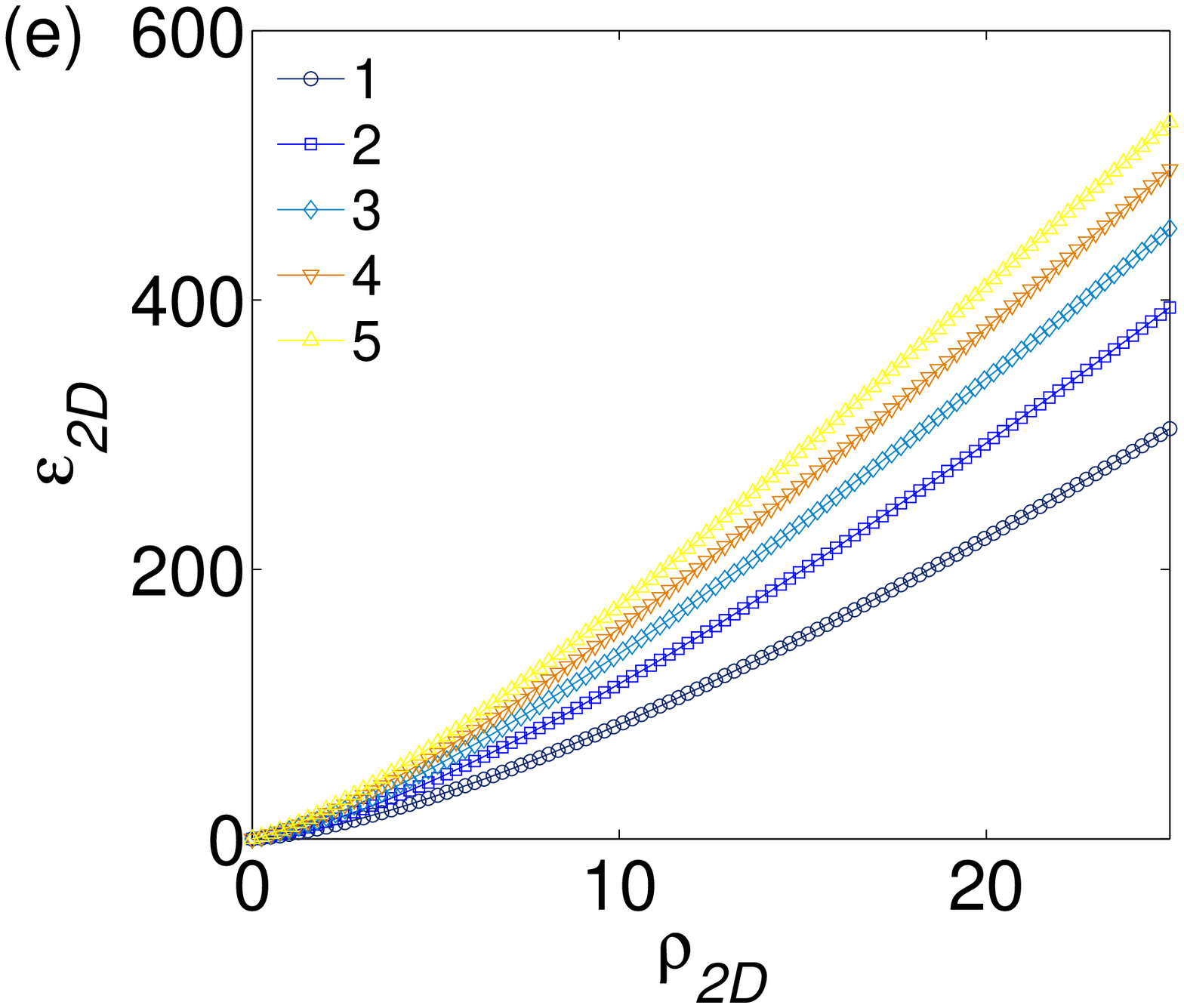}}
\resizebox{\textwidth}{!}{
\includegraphics{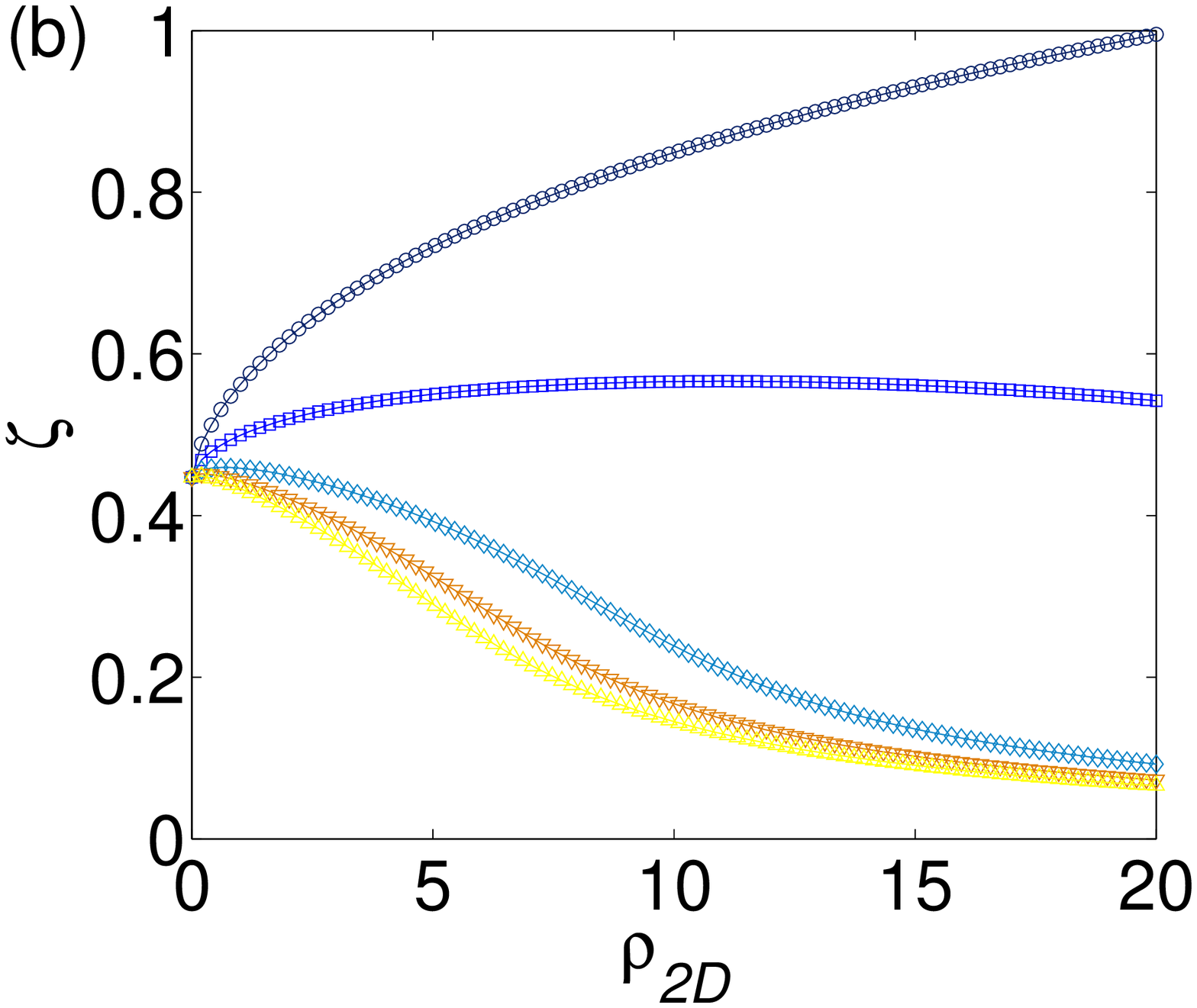}
\includegraphics{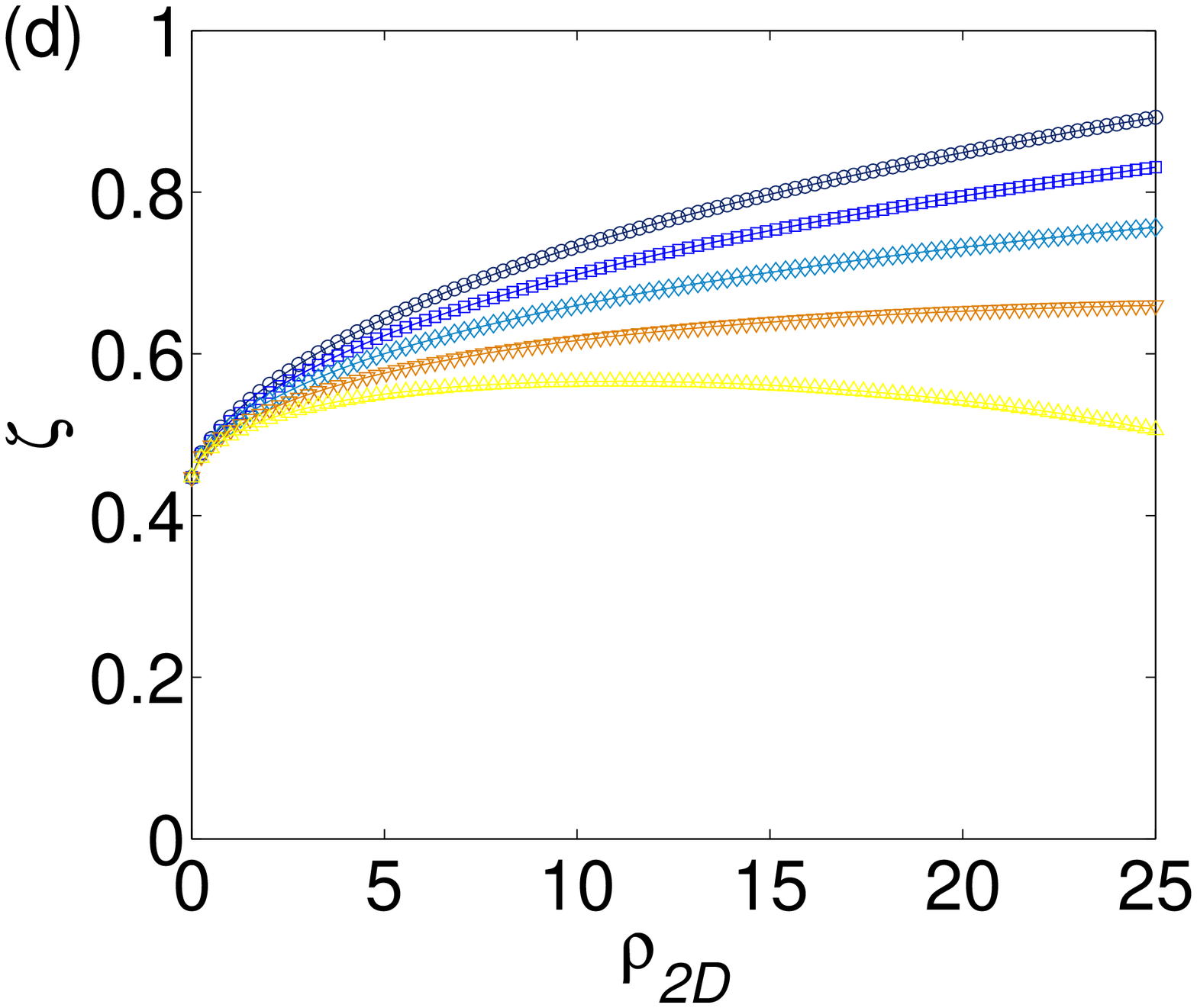}
\includegraphics{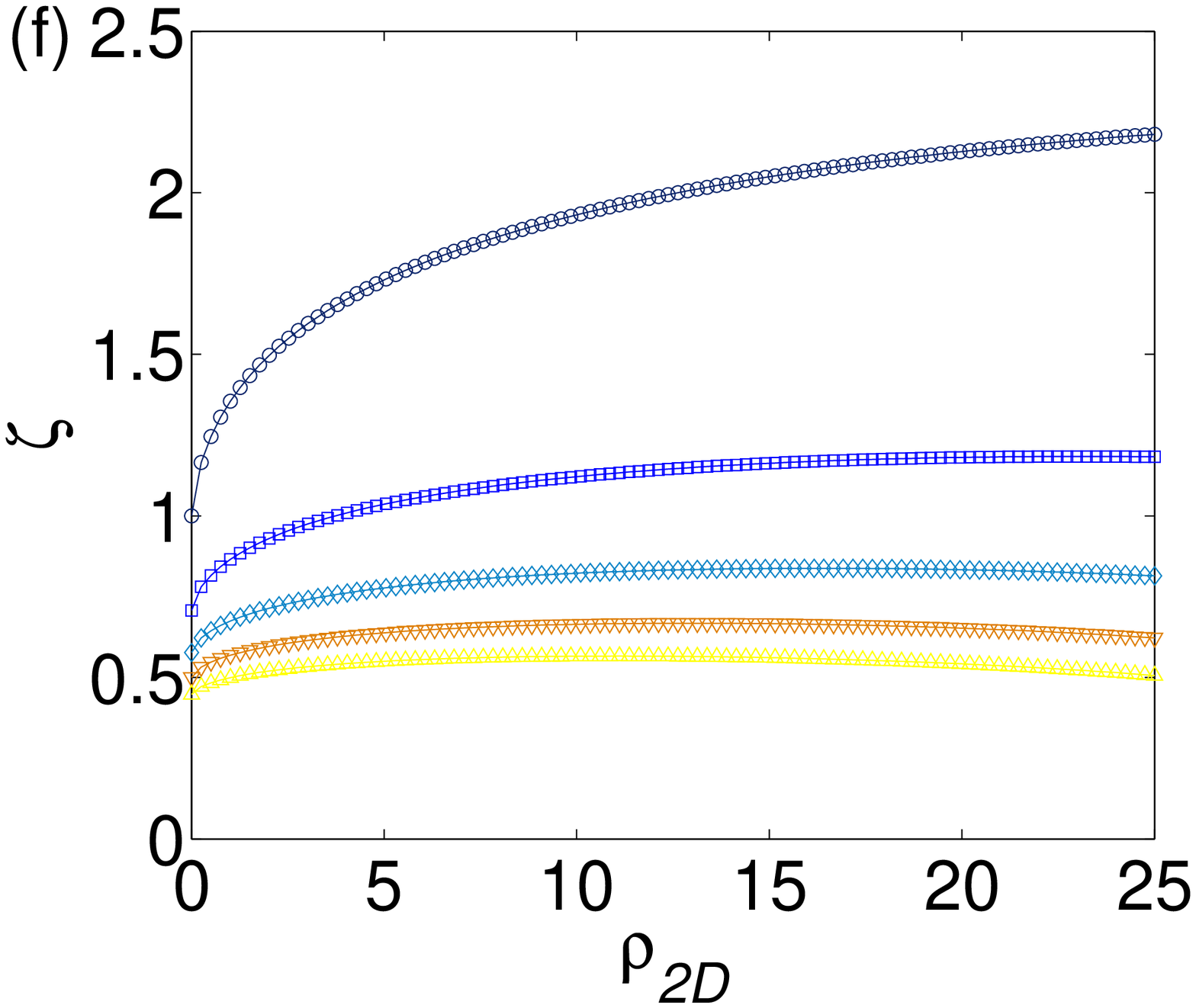}} 
\caption{(Color online) The energy density and transverse size, ${{\protect%
\varepsilon }_{\mathrm{2D}}}$ and $\protect\zeta $, versus the 2D
atomic density, $\protect\rho _{\mathrm{2D}}$. (a) \ and (b): The
dependences are shown for five different values of the spin,
$s=0,1/2,3/2,5/2$ and $7/2$, at the fixed negative scattering
length, ${a_{s}}/a_{c}=-0.2$, and fixed confinement strength,
${\protect\varpi }_{z}=5$. (c) and (d): The same for
five different values of the scattering length, $%
a_{s}/a_{c}=0,-0.05,-0.10,-0.15$ and $-0.20$, while $s=1/2$ and ${\protect%
\varpi }_{z}=5$ are fixed. (e) and (f): For five different values of the
transverse-confinement strength, ${{\protect\varpi }_{z}}=1,2,3,4$ and $5$,
at fixed $s=1/2$ and ${a_{s}}/a_{c}=-0.2$. The optical lattice is absent in
all the cases.}
\label{fig:1}
\end{figure}

The plots in Fig.~\ref{fig:1} display the dependence of the energy density $%
\varepsilon _{\mathrm{2D}}$ and transverse width $\zeta $ on the 2D density $%
\rho _{\mathrm{2D}}$ for the states uniform in the $\left( x,y\right) $
plane, as produced by Eqs.~(\ref{E-13}) and Eq.~(\ref{E-15}), respectively
(recall that ${\rho _{\mathrm{2D}}}\equiv {\left\vert \vartheta \right\vert
^{2}}$). In particular, Figs.~\ref{fig:1}(a) and ~\ref{fig:1}(b) show how
these dependences are affected by the fermion's spin, $s$. In Fig.~\ref%
{fig:1}(a), the energy density of the polarized gas, with $s=0$, is a
growing function of the atomic density, since in this case Eq. (\ref{E-3})
gives $g=0$, i.e., the Pauli exclusion principle suppresses the interaction
between the fermions, as mentioned above. When the gas is composed of equal
numbers of atoms with spins up and down ($s=1/2$), the attraction ($a_{s}<0$%
) between atoms with opposite directions of the spin causes a decrease in
the energy density, in comparison to the fully polarized gas. For the cases
of $s=3/2,5/2$ and $7/2$, it is observed that a maximum of the energy
density is attained at some atomic density ($\rho _{\mathrm{2D},\max }$),
due to the greater contribution of the attractive interactions in comparison
with the quantum pressure. In agreement with this trend, $\rho _{\mathrm{2D}%
,\max }$ is seen to be smaller for higher values of $s$, as the greater
number of possible spin states allows each atom to directly interact with a
larger number of atoms in other spin states. Furthermore, Fig.~\ref{fig:1}%
(b) shows that (for the same parameters as in Fig.~\ref{fig:1}(a)) the gas
with $s=1/2$ slowly shrinks in the transverse direction ($\zeta $ gradually
decreases) with the increase of the 2D density, after attaining a weakly
pronounced maximum at certain value of $\rho _{\mathrm{2D}}$. It is also
observed that, at the vanishingly small density, the transverse size does
not depend of spin $s$, in agreement with Eq.~(\ref{E-17}). On the other
hand, Figs.~\ref{fig:1}(c) and ~\ref{fig:1}(d) show that, for $s=1/2$, the
energy density remains a growing function of the atomic density for all the
values of $a_{s}/a_{c}$ considered, and, naturally, the growth is steeper
for smaller absolute values of the negative scattering length. Further, the
curves in Fig.~\ref{fig:1}(e) show that the energy density grows steeper
when the confinement strength, $\varpi _{z}$, is higher, and, naturally, the
tighter confinement leads to the smaller transverse size, as seen in Fig.~%
\ref{fig:1}(f). In summary, for $s>1/2$ and in the range
of the particle density ($\rho _{\mathrm{2D}}$) considered here (see Fig.~%
\ref{fig:1}(a)), the energy density ($\varepsilon _{\mathrm{2D}}$)
is an increasing function of the particle density up to a certain
value of the density ($\rho =\rho _{\mathrm{2D,}\max }$), where
$\varepsilon _{\mathrm{2D}}$ attains a maximum, which it followed by
the decrease of the energy density. It is also observed that
attaining this maximum is always preceded by the gradual tightening
of the gas in the confined direction and increase of the particle
density. In dynamical simulations, the system suffers the collapse
if the initial particle density exceeds the maximum value $\rho _{\mathrm{2D,%
}\max }$ (in a nonuniform configurations, the particles are pulled towards
the point where this maximum is reached), i.e., $\rho _{\mathrm{2D,}\max }$
is a separatrix bordering between stable and unstable uniform states, cf.
similar results reported in Refs. \cite{Adhikari2007b,Adhikari2007c,Chen2006}%
.

\begin{figure}[tbp]
\centering
{\normalsize
\resizebox{0.8\textwidth}{!}{\includegraphics{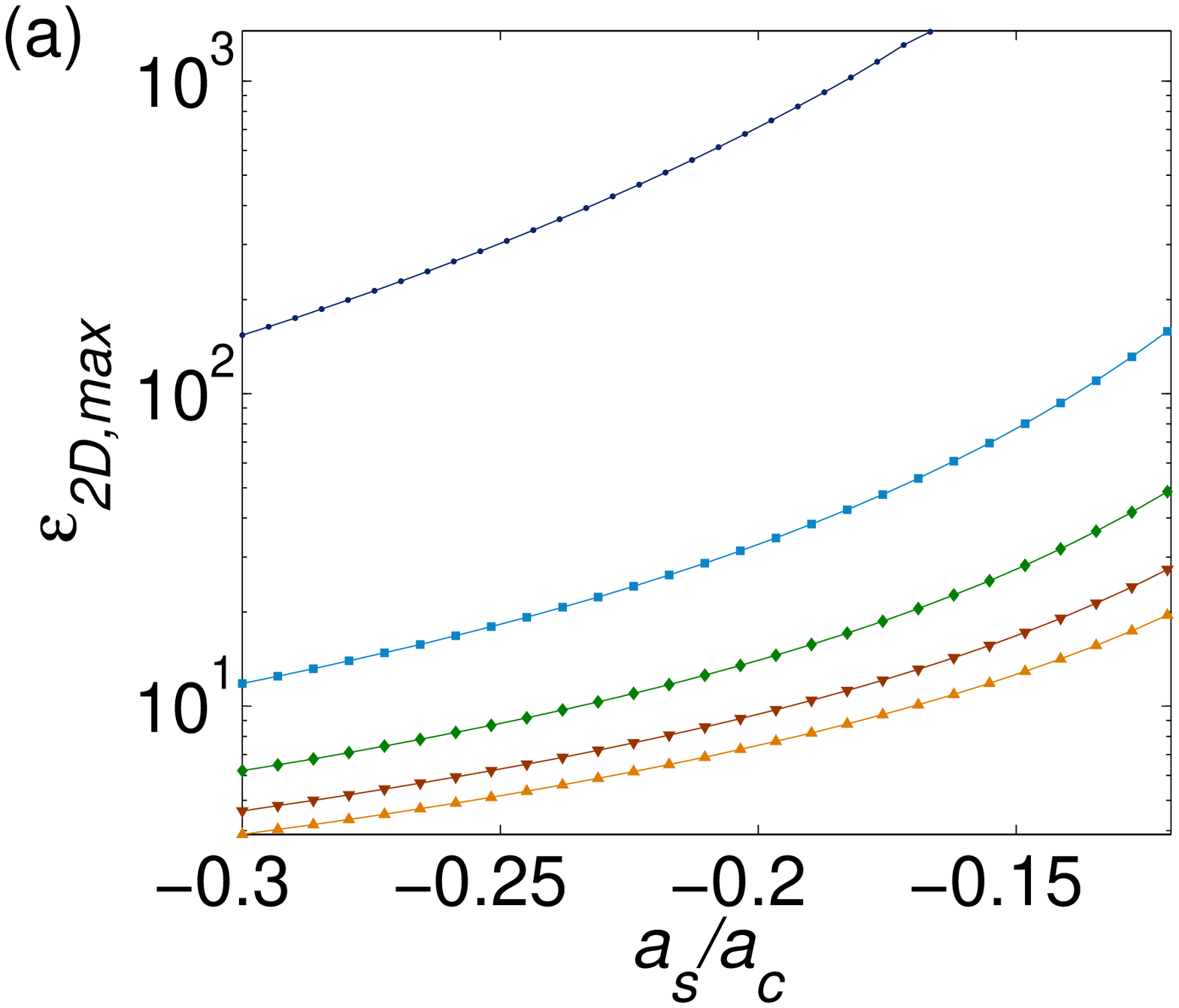}
\includegraphics{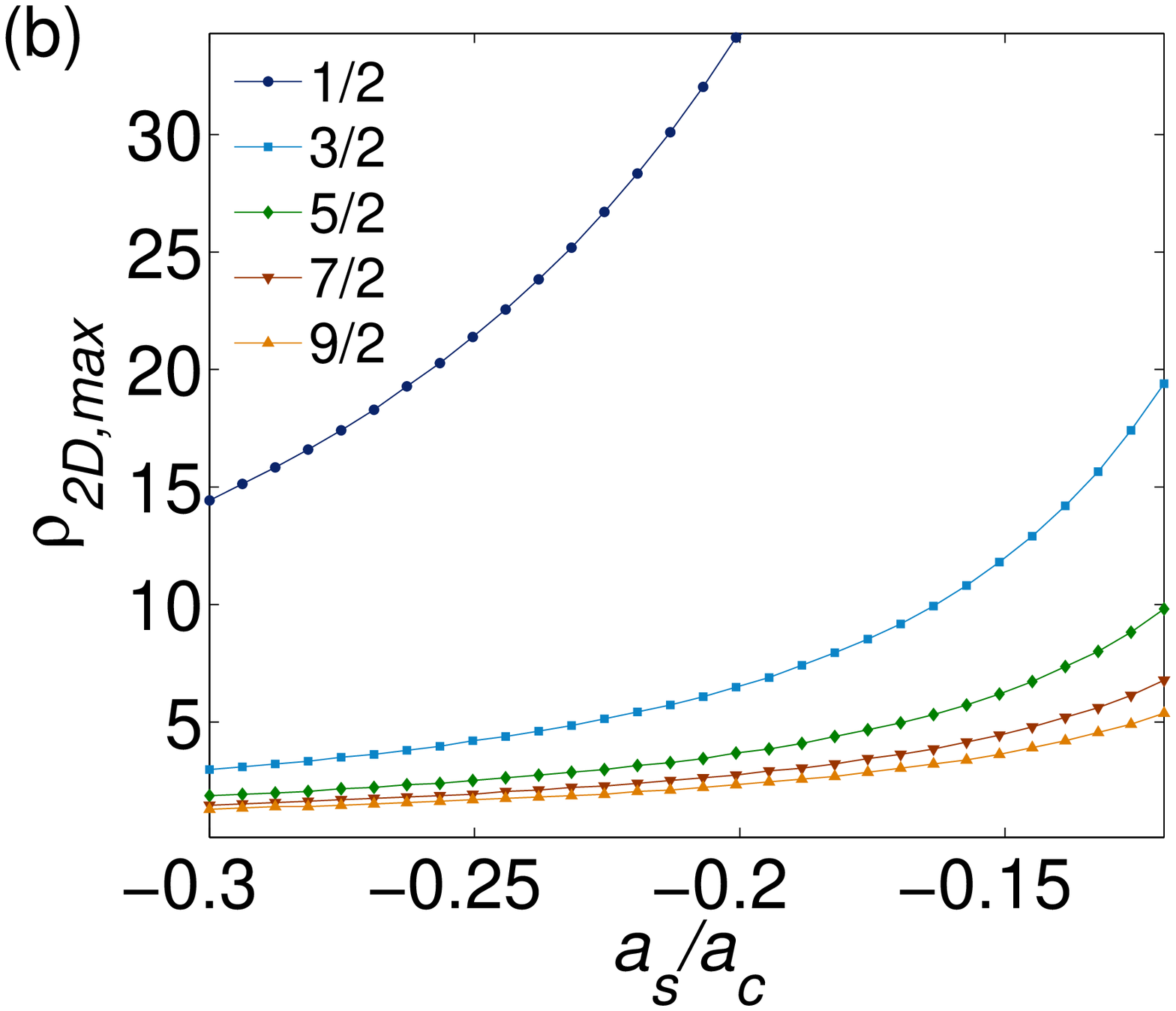}}  }
\caption{(Color online) (a) The maximum of the 2D energy density, $\protect%
\varepsilon _{\mathrm{2D},\max }$, as a function of the negative
scattering
length, $a_{s}/a_{c}$. (b) The 2D atomic density, $\protect\rho _{\mathrm{2D}%
,\max }$, at which $\protect\varepsilon _{\mathrm{2D}}$ attains its
maximum. The curves in both plots correspond to five different
values of spin $s$, as
indicated in the inset in (b). The confinement strength is fixed to be $%
\protect\varpi _{z}=5$, and the optical lattice is absent here.}
\label{fig:2}
\end{figure}

Figure~\ref{fig:2} shows the maximum value of the energy density as a
function of the strength of the attraction between the fermions in the
different spin states. It is seen that the enhancement of the attraction
causes a transition to lower energy and atomic densities. This behavior is
stronger pronounced for lower values of the spin, as a result of the larger
relative strength of the Pauli repulsion, leading to greater variations in
the energy-density maximum. It should be stressed, though, that these
results are approximate due to limitations imposed by the BCS regime
(overall, the weak interaction). Actually, most calculations reported in
this work were performed for the energy densities far from the maximum
value.

After the consideration of the spatially uniform states, the main objective
of our analysis is to identify the ground state of the fermion gas loaded
into the 2D potential consisting of the harmonic-oscillator trap and square
OL,

\begin{equation}
{U_{{\mathrm{2D}}}}=\varpi _{x}^{2}{X^{2}}+\varpi _{y}^{2}{Y^{2}}+{A_{x}}{%
\sin ^{2}}\left( {{\kappa _{x}}X}\right) +{A_{y}}{\sin ^{2}}\left( {{\kappa
_{y}}Y}\right) ,  \label{E-19}
\end{equation}%
where $\kappa _{x,y}\equiv 2\pi /\lambda _{x,y}$ are the lattice
wavenumbers. We start the presentation of the results for nearly-2D trapped
Fermi gas in the absence of the OL.

\subsection{The trapped quasi-2D gas in the absence of the optical lattice}

\begin{figure}[tbp]
\centering
{\small
\resizebox{0.5\textwidth}{!}{\includegraphics{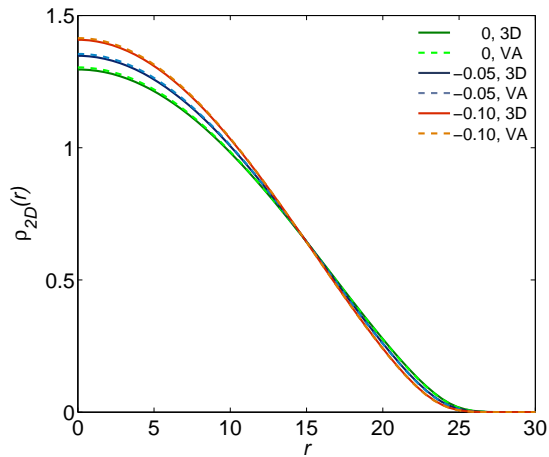}}  }
\caption{(Color online) The 2D radial density, $\protect\rho _{\mathrm{2D}%
}(r)$, as obtained from the full 3D equation, and from the 2D
reduction based on the variational approximation (``VA") for the
weak
isotropic harmonic-oscillator trap acting in the 2D plane ($\protect\varpi %
_{x}=\protect\varpi _{y}=0.1$, $\protect\varpi _{z}=2$), and $N=1000$.
Different curves correspond to the indicated values of $a_{s}/a_{c}=0,-0.05$
and $-0.1$. The optical-lattice potential is absent in this case.}
\label{fig:3}
\end{figure}

First, we have verified the accuracy of the 2D reduction by comparing
results generated by this approximation versus those obtained by integrating
the 3D equation (\ref{E-2}). The ground state was found by means of the
imaginary-time integration based on the fourth-order Runge-Kutta algorithm.
The comparison is produced in Fig.~\ref{fig:3}, where the radial-density
profiles are plotted for the isotropic 2D harmonic-oscillator trap ($\varpi
_{x}=\varpi _{y}$), showing an excellent agreement between the 2D and full
3D descriptions. Actually, this is one of main results of the present work.

\subsection{Effects of the optical lattice}

\begin{figure}[tbp]
\centering
{\normalsize
\resizebox{0.8\textwidth}{!}{\includegraphics{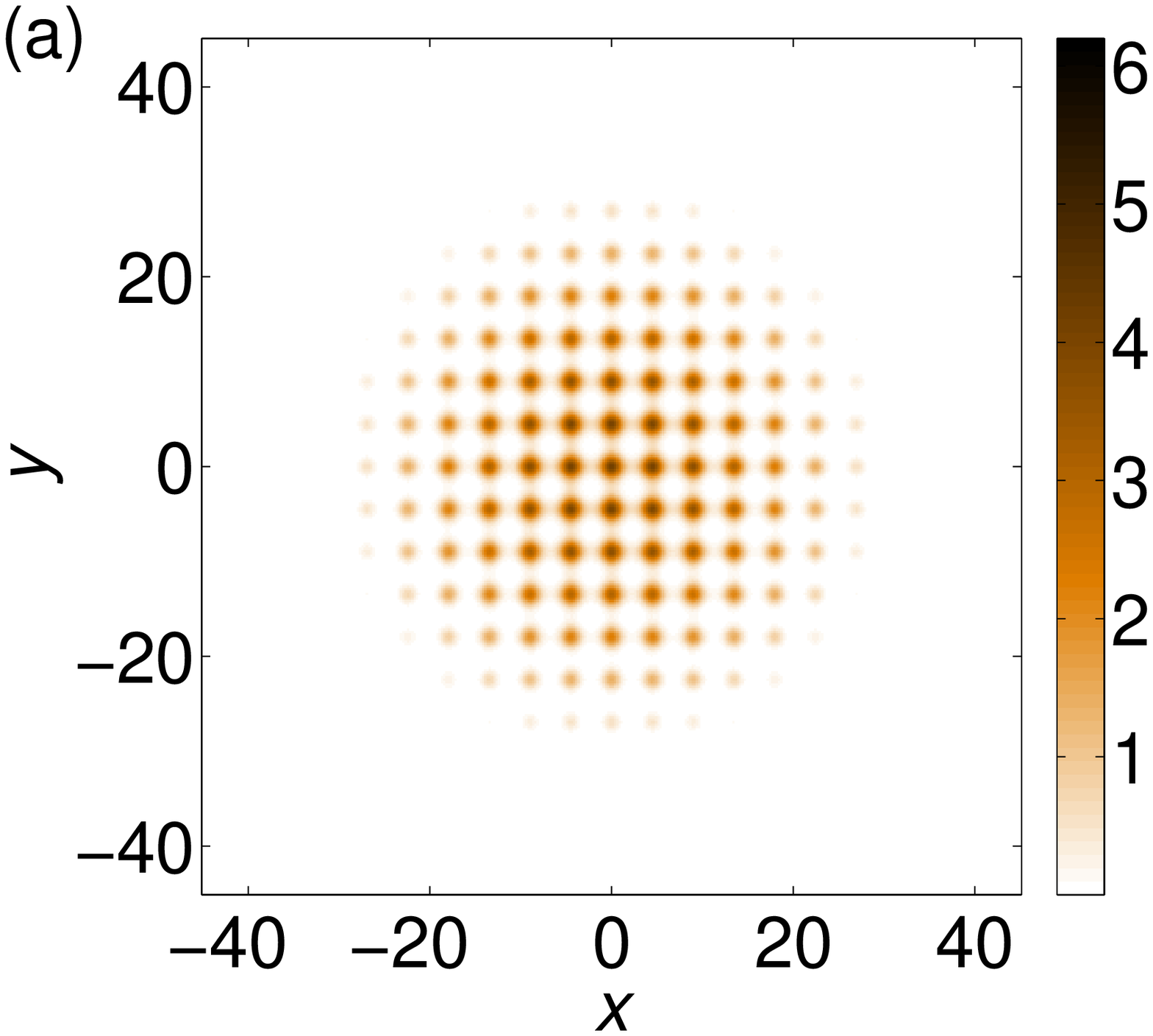}
\includegraphics{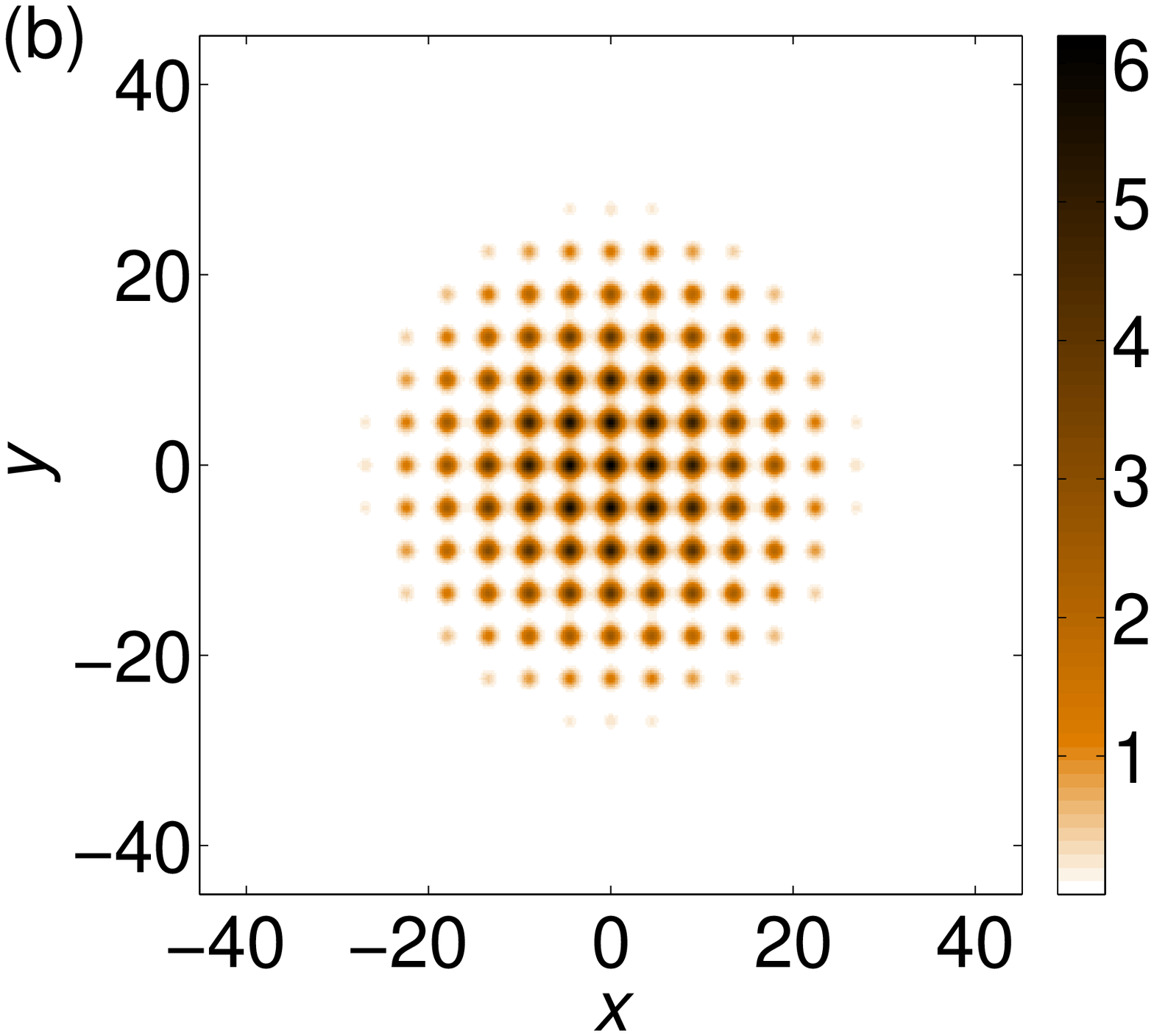}
}
\resizebox{0.8\textwidth}{!}{\includegraphics{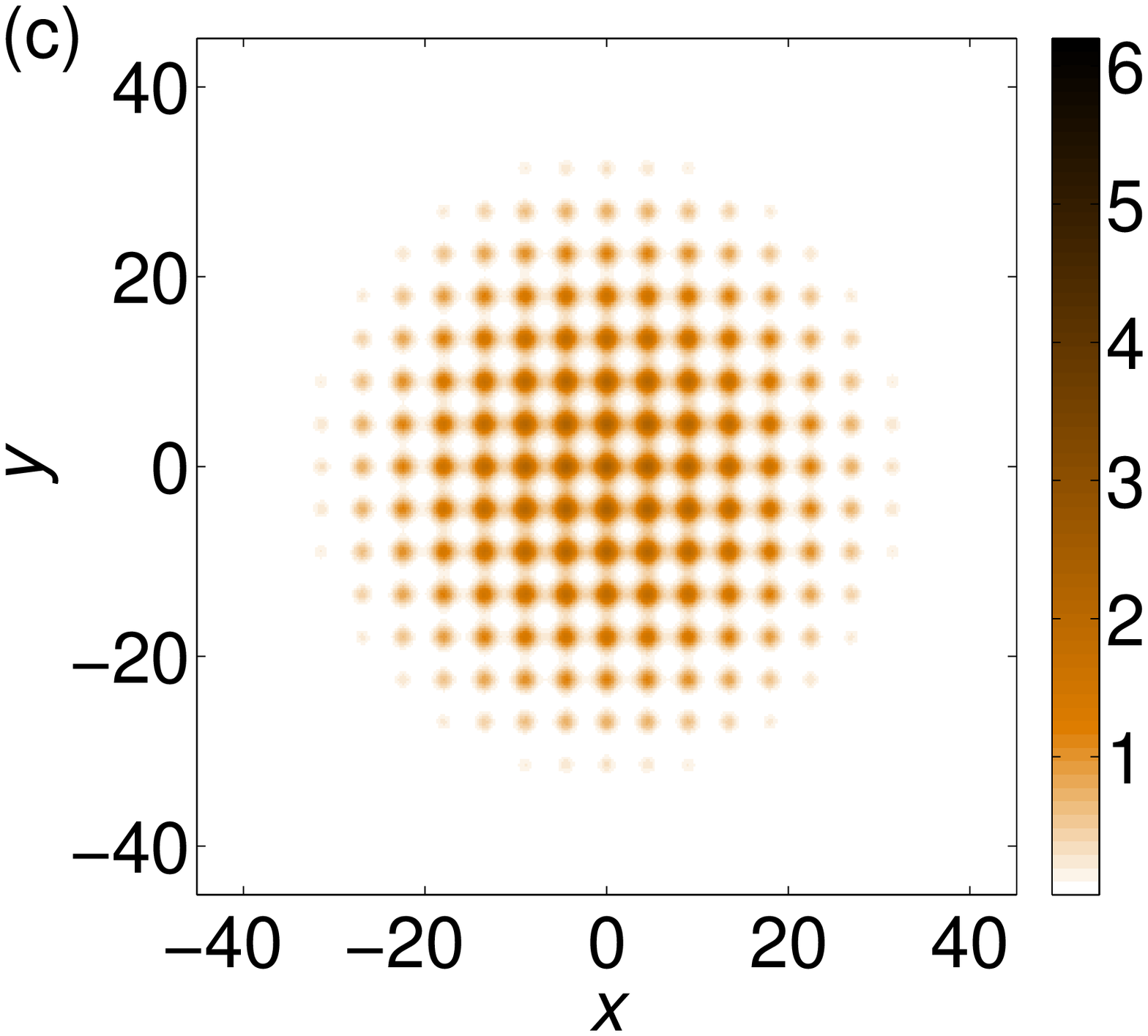}
\includegraphics{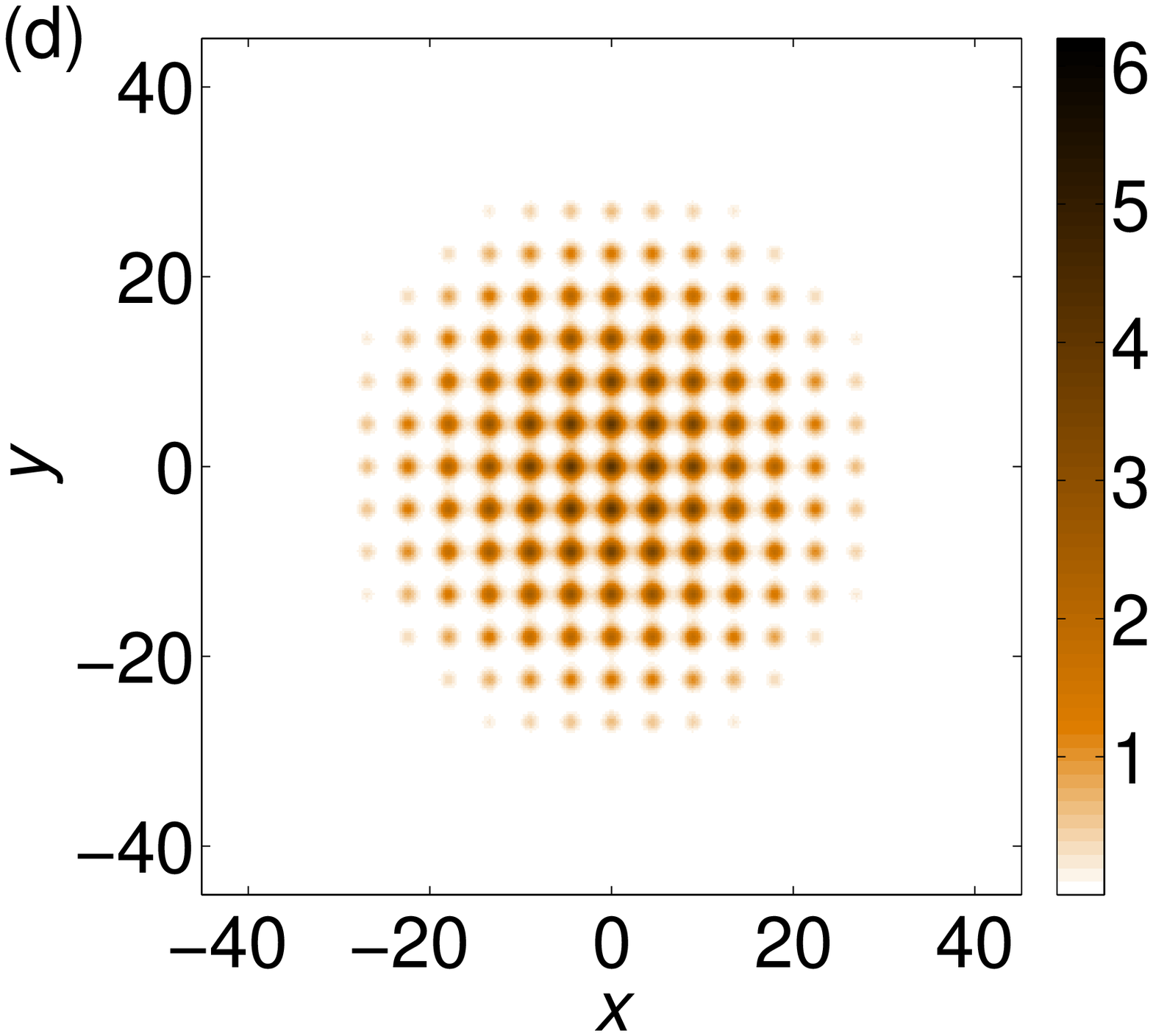}
}  }
\caption{(Color online) Results for the quasi-2D gas trapped in the
combination of the optical lattice and weak harmonic-oscillator potential.
The plots display density profiles, $\protect\rho _{\mathrm{2D}}(x,y)$, for
four cases: (a) $a_{s}/a_{c}=0$ and $\protect\varpi _{z}=2$, (b) $%
a_{s}/a_{c}=-0.2$ and $\protect\varpi _{z}=2$, (c) $a_{s}/a_{c}=0$ and $%
\protect\varpi _{z}=8$, and (d) $a_{s}/a_{c}=-0.2$ and $\protect\varpi _{z}=8
$. The remaining parameters are $N=1000$, $\protect\varpi _{x}=\protect%
\varpi _{y}=0.1$, $A_{x}=A_{y}=10$, and $\protect\lambda _{x}=\protect%
\lambda _{y}=9$.}
\label{fig:4}
\end{figure}

\begin{figure}[tbp]
\centering
{\normalsize
\resizebox{\textwidth}{!}{
\includegraphics{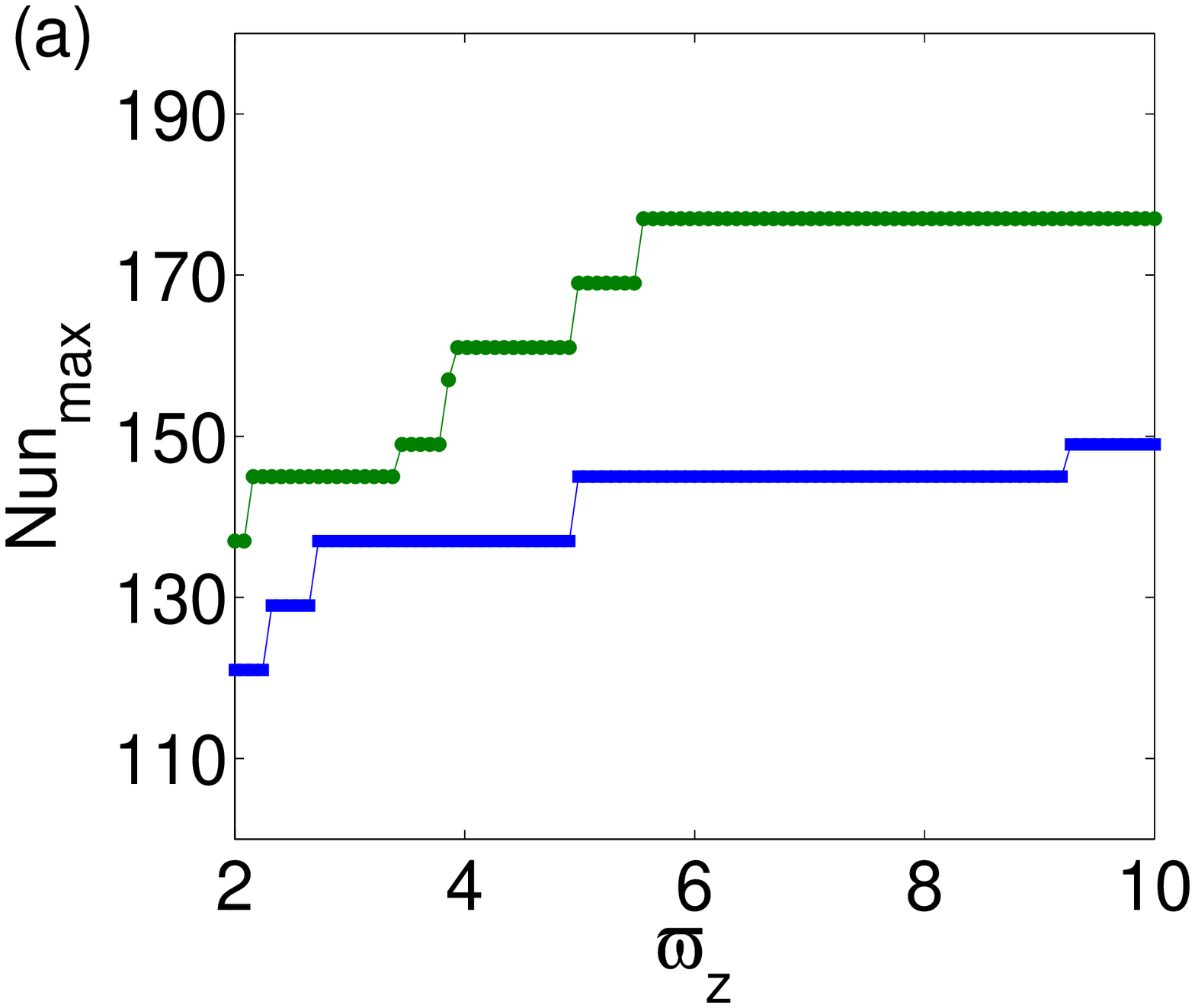}
\includegraphics{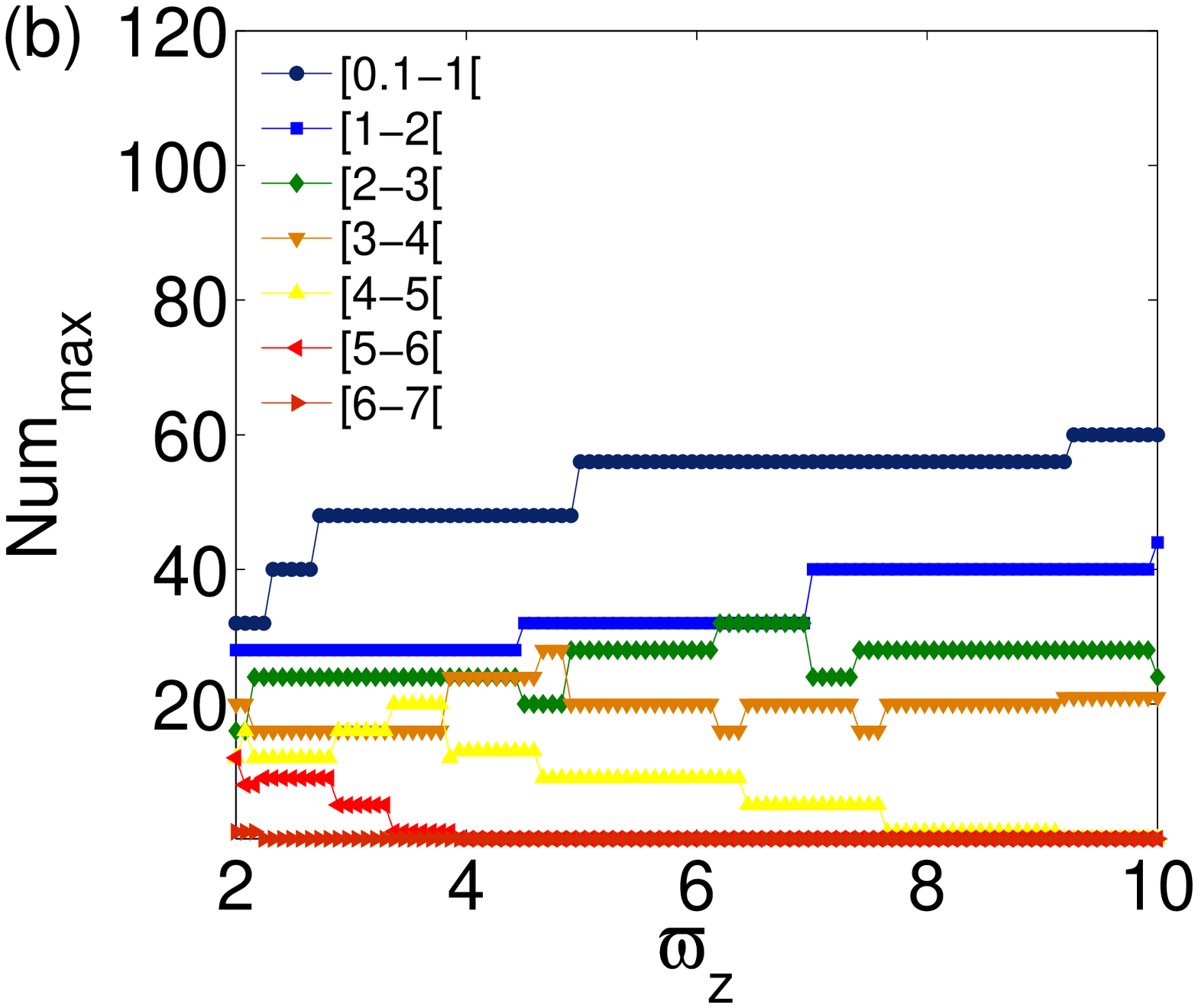}
\includegraphics{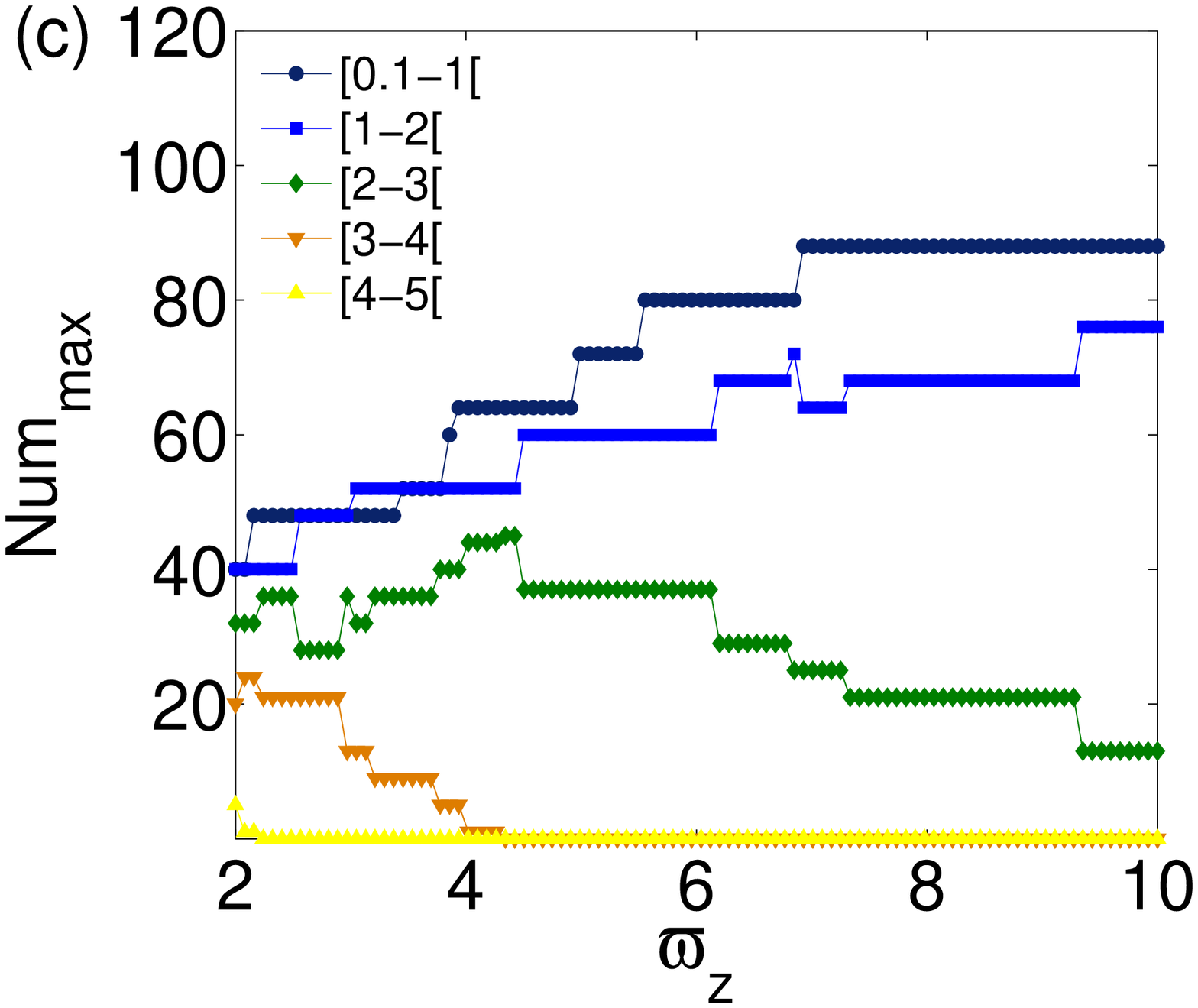}
}  }
\caption{(Color online) (a) The number of density peaks, located at minima
of the OL potential, as a function of confinement strength $\protect\varpi %
_{z}$. The area occupied by the gas is defined as that with $\protect\rho _{%
\mathrm{2D}}\geq 0.1$. The blue and green lines correspond to $%
a_{s}/a_{c}=-0.2$ and $a_{s}/a_{z}=0$, respectively. Panels (b) and (c)
display, severally, detailed pictures for $a_{s}/a_{c}=-0.2$ and $%
a_{s}/a_{c}=0$, in seven different ranges of density$\protect\rho _{\mathrm{%
2D}}$, as indicated in the inset. The parameters are $N=1000$, $\protect%
\varpi _{x}=\protect\varpi _{y}=0.1$, $A_{x}=A_{y}=10$, and $\protect\lambda %
_{x}=\protect\lambda _{y}=9$.}
\label{fig:5}
\end{figure}

The plots in Fig.~\ref{fig:4} show the density distribution in the
ground state of the gas loaded into the 2D harmonic-oscillator
potential combined with the 2D square OL (see Eq.~\ref{E-19}).
Naturally, density peaks
coincide with local minima of the OL potential. Figures~\ref{fig:4}(a) and %
\ref{fig:4}(b) demonstrate that the attractive interaction leads to a
decrease in the number of peaks visible in the density profile, and to an
increase in their heights. On the other hand, Figs.~\ref{fig:4}(c) and ~\ref%
{fig:4}(d) show that, for the same interaction strength as above,
tighter transverse confinement results in a greater number of peaks,
with a lower atomic density at them. Due to the symmetry of the OL,
the number of visible peaks must be in multiples of $4$. We define
as ``observable" an area with density $\rho _{\mathrm{2D}}\geq 0.1$.
The plot of Fig.~\ref{fig:5}(a) shows the dependence of the
observable number of peaks on the transverse-confinement strength
$\varpi _{z}$, for $a_{s}/a_{z}=0$
(green circles) and $a_{s}/a_{z}=-0.2$ (blue squares). At all values of $%
\varpi _{z}$, the number of visible peaks is higher in the $a_{s}=0$ case,
with a clear nonlinear dependence that is characterized by intervals in the
scale of in $\varpi _{z}$ corresponding to particular numbers of the peaks.
This feature is particularly noticeable in the range of $\varpi _{z}=[5,10]$.

Figures~\ref{fig:5}(b) and ~\ref{fig:5}(c) show detail of the peaks
characterization, which amounted to counting the number of peaks in seven
specific zones of values of $\rho _{\mathrm{2D}}$: [0.1,1[, [1,2[, [2,3[,
[3,4[, [4,5[, [5,6[ and [6,7[. The findings uncover intricate changes in the
density patterns, following variations of the system's parameters. In
particular, Fig.~\ref{fig:5}(b) shows that, for $a_{s}/a_{z}=-0.2$, distinct
zones of the values of $\varpi _{z}$ may cluster at the same number of
peaks, suggesting that a given number of peaks may be realized in various
configurations. The plot in Fig.~\ref{fig:5}(c) shows the non-interacting
case, $a_{s}=0$, with no visible peak in the last two ranges, which implies
greater expansion of the gas in the 2D plane. In each range, the variation
of parameters significantly alters the distribution of peaks. The multitude
of the different coexisting robust multi-peak patterns suggests that this
setting has a potential for the use as a data-storage system.

\begin{figure}[tbp]
\centering
{\normalsize
\resizebox{0.8\textwidth}{!}{\includegraphics{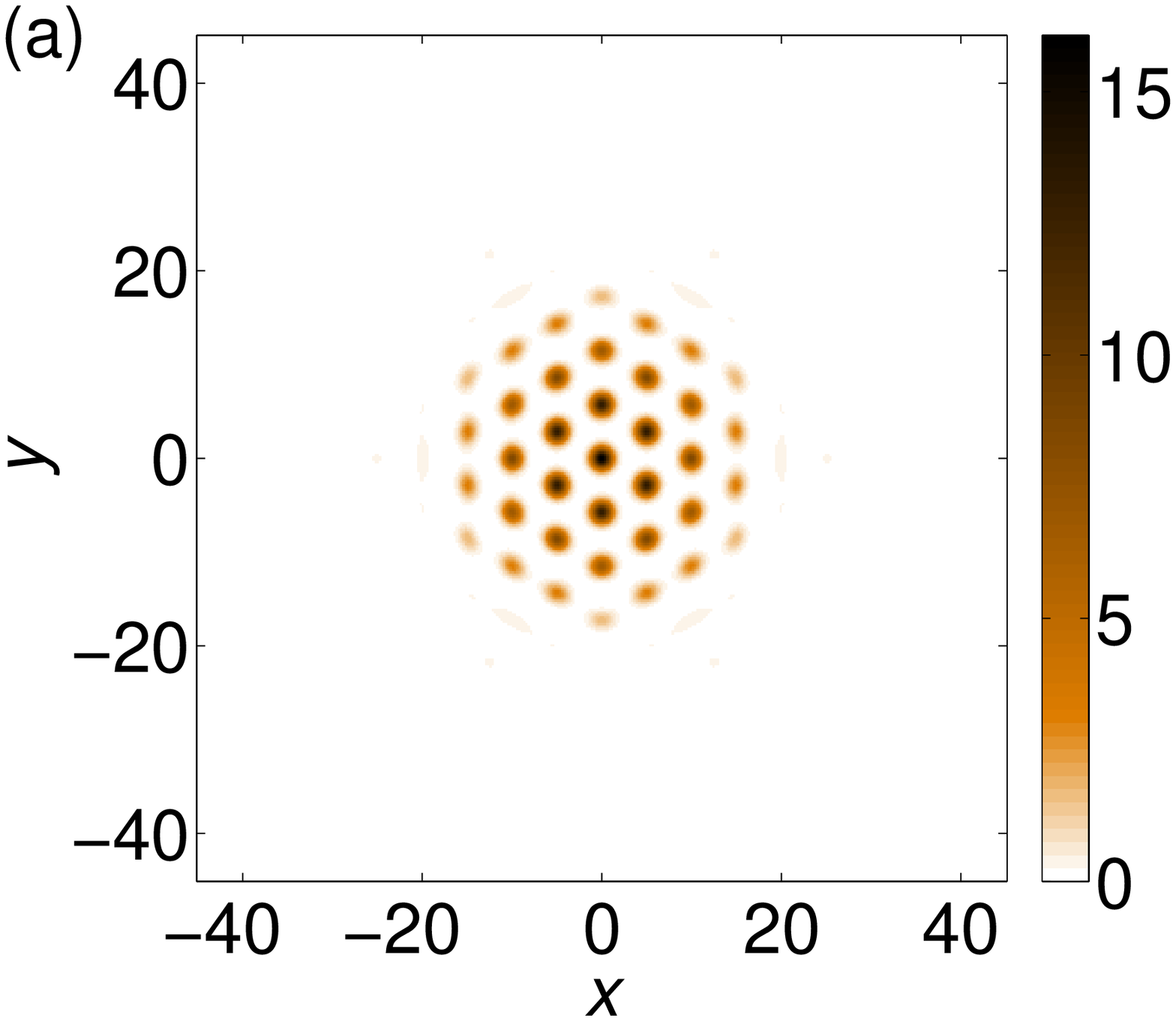}
\includegraphics{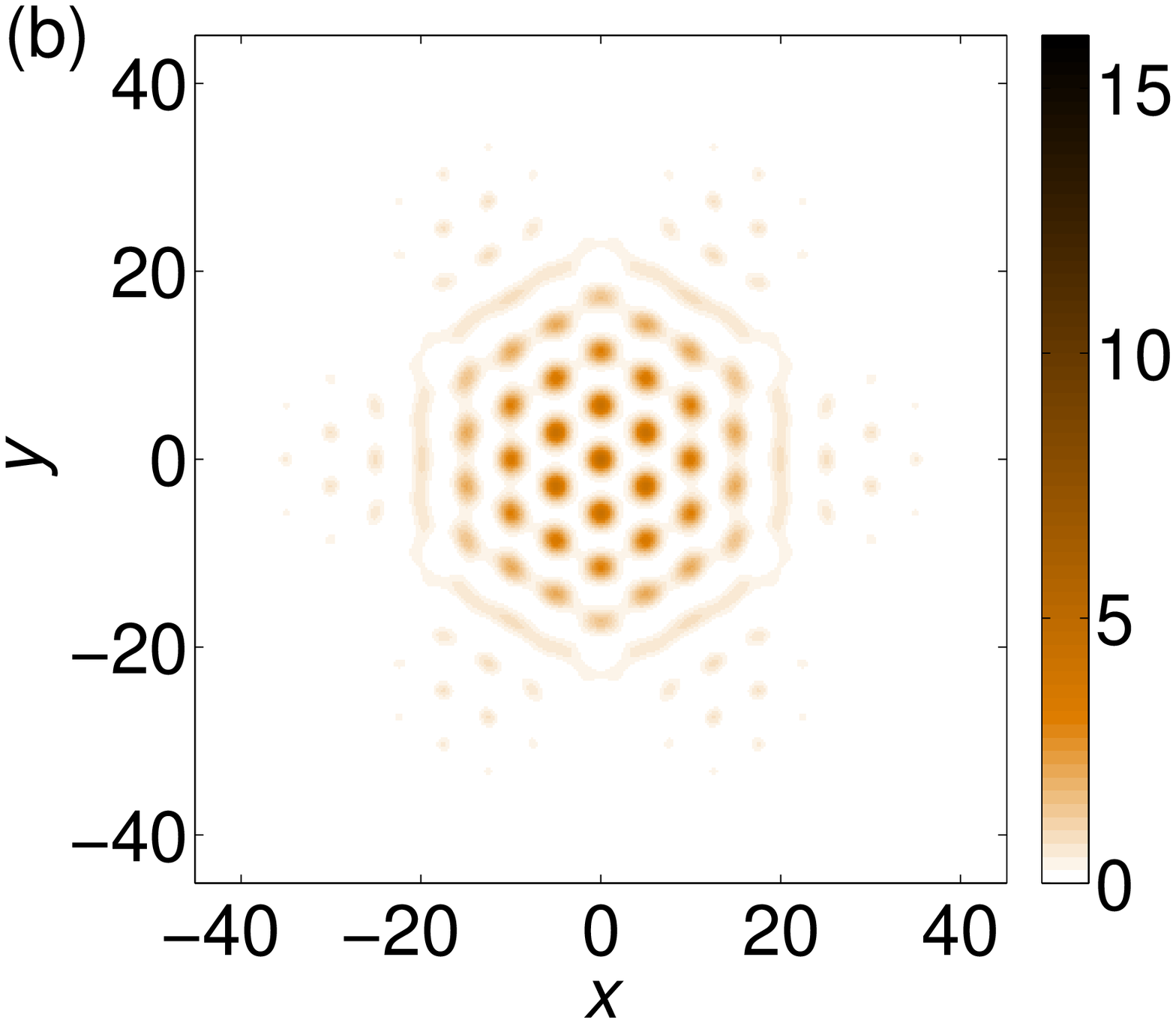}
}  }
\caption{(Color online) The same as in Fig. \protect\ref{fig:4} for the
superlattice built of two triangular OLs with angle $\protect\theta %
=8^{\circ }$ between them: (a) $a_{s}/a_{z}=-0.2$, $\protect\varpi _{z}=2$;
(b) $a_{s}/a_{z}=0$, $\protect\varpi _{z}=5$. The other parameters are $%
N=1000$, $\protect\varpi _{x}=\protect\varpi _{y}=0.1$, and $A_{T}=5$, $%
\protect\lambda _{T}=10$.}
\label{fig:6}
\end{figure}

Figure~\ref{fig:6} displays the results for the Fermi gas in the presence of
a superlattice formed by a superposition of two triangular lattice, each
formed by the set of three pairs of counter-propagating laser-beams, with
angles $60^{\circ }$ between them, and equal amplitudes and wavelengths. To
build the superlattice, we fix the first pair of beams in the $x$-direction
in both triangular lattices, and then rotate the two lattices clockwise and
counterclockwise by $\pm 4^{\circ }$. The overall shape of the thus formed
superlattice remains hexagonal. It is observed that the density maxima are
most visible within the central hexagon. The set of different patterns
trapped in the superlattice is richer than in the case of the square OL
considered above, as different potential minima of the superlattice
potential have different depths.

\section{The one-dimensional reduction}

\label{sec:3}

\subsection{The derivation of the 1D equations}

To perform the reduction of the mean-field description from 3D to 1D, we
consider an external potential formed by the harmonic-oscillator potential
in the $\left( x,y\right) $-plane, which represents the cylindrically
symmetric magnetic confinement, and an axial ($z$-dependent) potential:
\begin{equation}
V\left( {{\mathbf{r}},t}\right) =\frac{1}{2}m\omega _{\mathrm{t}%
}^{2}r^{2}+V_{\mathrm{1D}}\left( {z,t}\right) ,  \label{E-20}
\end{equation}%
where $r\equiv \sqrt{x^{2}+y^{2}}$. Again, the initial ansatz assumes the
factorization of the 3D wave function into a product of $r$- and $z$%
-dependent functions. The former one is taken as the ground state of the 2D
harmonic oscillator, hence the ansatz is

\begin{equation}
\Psi \left( {{\mathbf{r}},t}\right) =\frac{1}{\sqrt{{\pi }}{\sigma \left( {%
z,t}\right) }}{\exp }\left( -\frac{r^{2}}{{2\sigma }^{2}{{{\left( {z,t}%
\right) }}}}\right) f\left( {z,t}\right) ,  \label{E-21}
\end{equation}%
where the 1D wave function, $f\left( z,t\right) $, is normalized to the
number of atoms, $N$, and the axial density is defined as $n_{\mathrm{1D}%
}\equiv \left\vert f\right\vert ^{2}$. Variational functions $f$ and $\sigma
$ can be determined by the minimization of the 3D action after performing
the integration in the $\left( x,y\right) $ plane, cf. the derivation of the
1D NPSE for the bosonic gas \cite{Salasnich2002}. This procedure gives rise
to the effective 1D Lagrangian density,
\begin{eqnarray}
{\mathcal{L}_{\mathrm{1D}}} &=&i\frac{\hbar }{2}\left( {{{f^{\ast }}}{%
\partial _{t}}{f}-{f}{\partial _{t}}{{f^{\ast }}}}\right) -\frac{{{\hbar ^{2}%
}}}{{2{m}}}{\left\vert {{\partial _{z}}{f}}\right\vert ^{2}}-{V_{\mathrm{1D}%
}n_{\mathrm{1D}}}  \nonumber \\
&&-\left[ \frac{{{\hbar ^{2}}}}{{2{m}}}\frac{3}{{5\sigma ^{4/3}}}C_{\mathrm{%
1D}}n_{\mathrm{1D}}^{5/3}+\frac{{g}}{{4\pi \sigma ^{2}}}n_{\mathrm{1D}}^{2}+%
\frac{{\hbar ^{2}}}{{2m{\sigma ^{2}}}}{n_{{\mathrm{1D}}}}+\frac{1}{2}m\omega
_{t}^{2}{\sigma ^{2}}{n_{{\mathrm{1D}}}}\right] ,  \label{E-22}
\end{eqnarray}%
where $C_{\mathrm{1D}}\equiv (3/5)(6\pi /(2s+1))^{2/3}$, $s$ is the
fermionic spin, as before, and the last two terms are the result of
integration, which keeps the 3D characteristics of the underlying system:
When the gas is homogeneous and the external potential is absent, the latter
term is irrelevant.

Similar to the 2D case, the normalization is performed with characteristic
length scale $a_{c}$ and frequency $\omega _{c}$, so that $Z\equiv z/a_{c}$,
$\psi \equiv a_{c}^{1/2}f$, $U_{\mathrm{1D}}\equiv 2V_{\mathrm{1D}}/(\hbar
\omega _{c})$, $G_{\mathrm{1D}}\equiv 8sa_{s}/(2s+1)a_{c}$, $\eta \equiv
\sigma /a_{c}$, and $\varpi _{\mathrm{t}}\equiv \omega _{\mathrm{t}}/\omega
_{c}$. In addition, we define the dimensionless 1D atom density, $\rho _{%
\mathrm{1D}}\equiv \left\vert {\psi }\right\vert ^{2}$. Then, the normalized
Lagrangian density is
\begin{eqnarray}
{{\tilde{\mathcal{L}}}_{\mathrm{1D}}} &=&\frac{i}{2}({{\psi ^{\ast }}}{%
\partial _{\tau }}{\psi }-{\psi }{\partial _{\tau }}{{\psi ^{\ast }}})-{%
\left\vert {{\partial _{Z}}{\psi }}\right\vert ^{2}}-{U_{\mathrm{1D}}{\rho }%
_{\mathrm{1D}}}  \nonumber \\
&&-\left[ \frac{3}{{5{\eta ^{4/3}}}}{C_{{\mathrm{1D}}}}\rho _{1D}^{5/3}+%
\frac{1}{{2{\eta ^{2}}}}{G_{{\mathrm{1D}}}}\rho _{1D}^{2}+\varpi _{t}^{2}{%
\eta ^{2}}{\rho _{{\mathrm{1D}}}}+\frac{1}{{\eta ^{2}}}{\rho _{{\mathrm{1D}}}%
}\right] .  \label{E-23}
\end{eqnarray}%
The 1D energy density $\varepsilon _{\mathrm{1D}}$ for uniform
states is
\begin{equation}
\varepsilon _{\mathrm{1D}}=\frac{3}{{5{\eta
^{4/3}}}}{C_{{\mathrm{1D}}}}\rho _{1D}^{5/3}+\frac{1}{{2{\eta
^{2}}}}{G_{{\mathrm{1D}}}}\rho _{1D}^{2}+\varpi
_{\mathrm{t}}^{2}{\eta ^{2}}{\rho _{{\mathrm{1D}}}}+\frac{1}{{\eta ^{2}}}{%
\rho _{{\mathrm{1D}}}},  \label{E-24}
\end{equation}%
cf. Eq. (\ref{E-13}) in the 2D case, which makes it possible to compare
contributions of the different interactions involved. In fact, the
comparison is not straightforward, because of the dependence of $\eta $ on $%
\rho _{\mathrm{1D}}$; however, for very low densities $\eta $ may be assumed
constant, in which case the last two terms dominate in $\varepsilon _{\mathrm{1D%
}}$. Now, varying the action with respect to $\psi $, we derive the
respective Euler-Lagrange equation (similar to the Eq.~\ref{E-5}):

\begin{equation}
i{\partial _{\tau }}{\psi }=\left[ {-\partial _{Z}^{2}+{U}_{\mathrm{1D}}}+%
\frac{{C_{\mathrm{1D}}}}{{\eta ^{4/3}}}{\left\vert {\psi }\right\vert ^{4/3}}%
+\frac{{G_{\mathrm{1D}}}}{{\eta ^{2}}}{\left\vert {\psi }\right\vert ^{2}}+%
\frac{1}{{\eta ^{2}}}+{\varpi _{\mathrm{t}}^{2}}\eta ^{2}\right]
{\psi }, \label{E-25}
\end{equation}

which is the 1D equation of motion, with powers of the transverse width, $%
\eta $, in the collisional and Pauli terms being different from their
counterparts in the 2D equation, cf. Eq. (\ref{E-14}). Further, varying the
action with respect to $\eta $ gives rise to the corresponding
Euler-Lagrange equation,

\begin{equation}
\varpi _{t}^{2}{\eta ^{4}}-\frac{2}{5}C_{\mathrm{1D}}{\left\vert \psi
\right\vert ^{4/3}}{\eta ^{2/3}}-\left( {1+\frac{G_{\mathrm{1D}}}{2}{{%
\left\vert \psi \right\vert }^{2}}}\right) =0,  \label{E-26}
\end{equation}%
which establishes the dependence of the transverse size, $\eta $, on density
$\rho _{\mathrm{1D}}$. Generally, this equation should be solved
numerically. In the low-density limit, it has a trivial solution, $\eta
^{2}=1/\varpi _{\mathrm{t}}$, which, if substituted in Eq.~(\ref{E-25})
allows one to derive a closed-form equation for $\psi $ [similar to the
situation for the same limit in the 2D case, cf. Eq. (\ref{E-18})]:

\begin{equation}
i{\partial _{\tau }}\psi =\left[ {-\partial _{Z}^{2}+{U}_{\mathrm{1D}}+C_{%
\mathrm{1D}}\varpi _{\mathrm{t}}^{2/3}{{\left\vert \psi \right\vert }^{4/3}}+%
{\varpi _{\mathrm{t}}}G_{\mathrm{1D}}{{\left\vert \psi \right\vert }^{2}}+2{%
\varpi }}_{t}\right] \psi ,  \label{E-27}
\end{equation}%
with the lowest-order nonlinear term accounting for the Pauli repulsion of
the same type as in the 2D equation (\ref{E-18}).

\begin{figure}[tbp]
\centering
{\normalsize
\resizebox{0.8\textwidth}{!}{\includegraphics{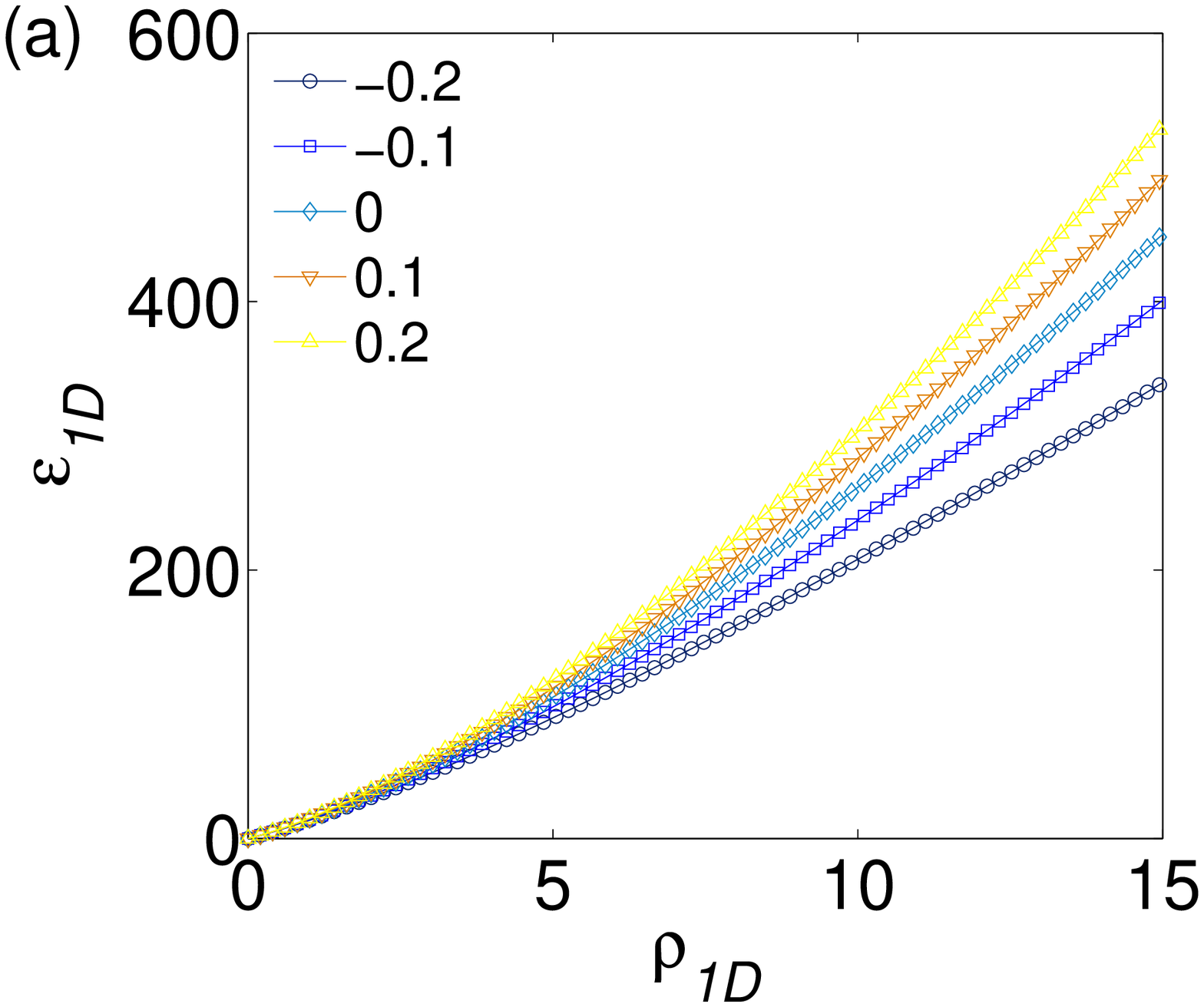}
\includegraphics{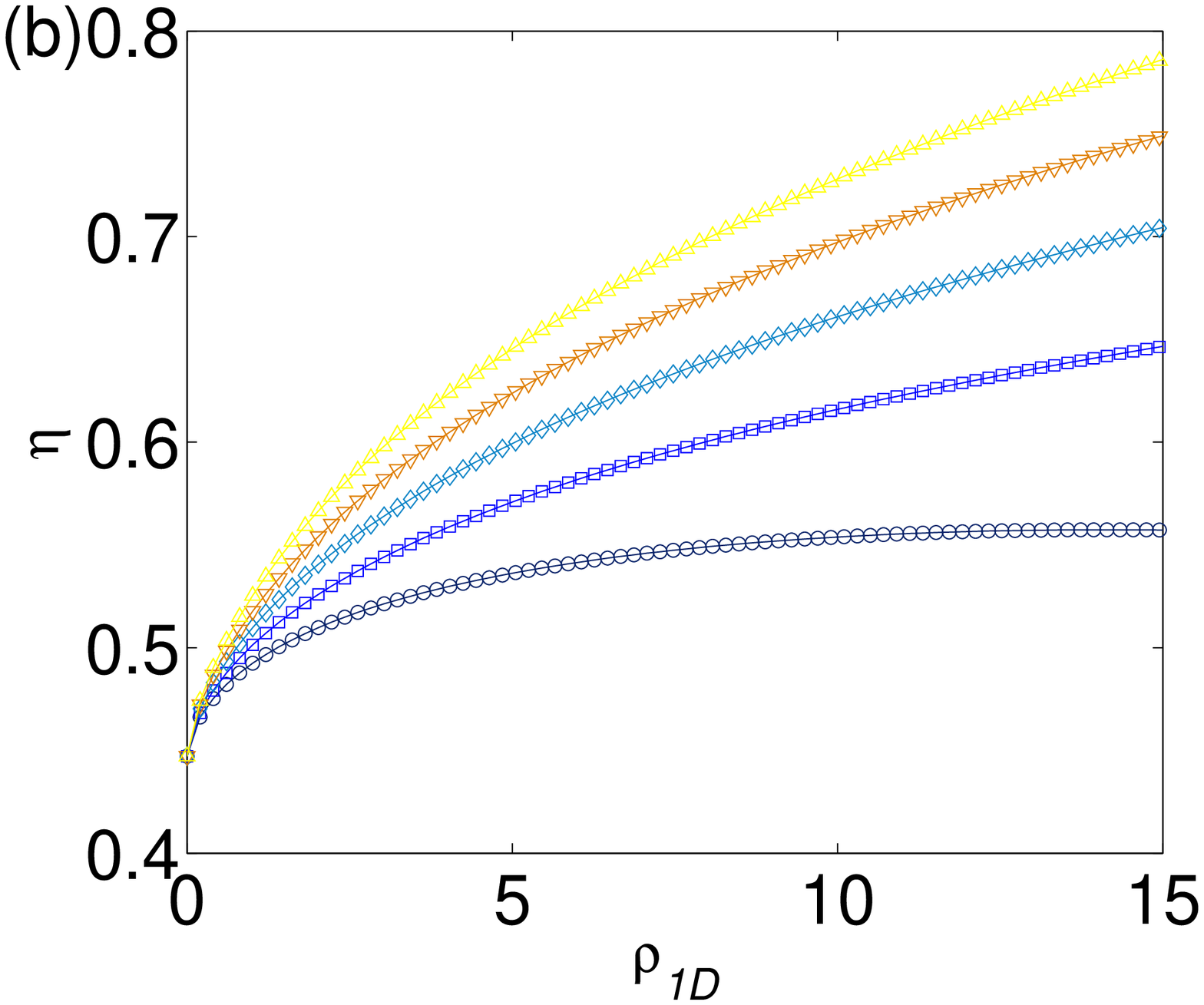}
}
\resizebox{0.8\textwidth}{!}{\includegraphics{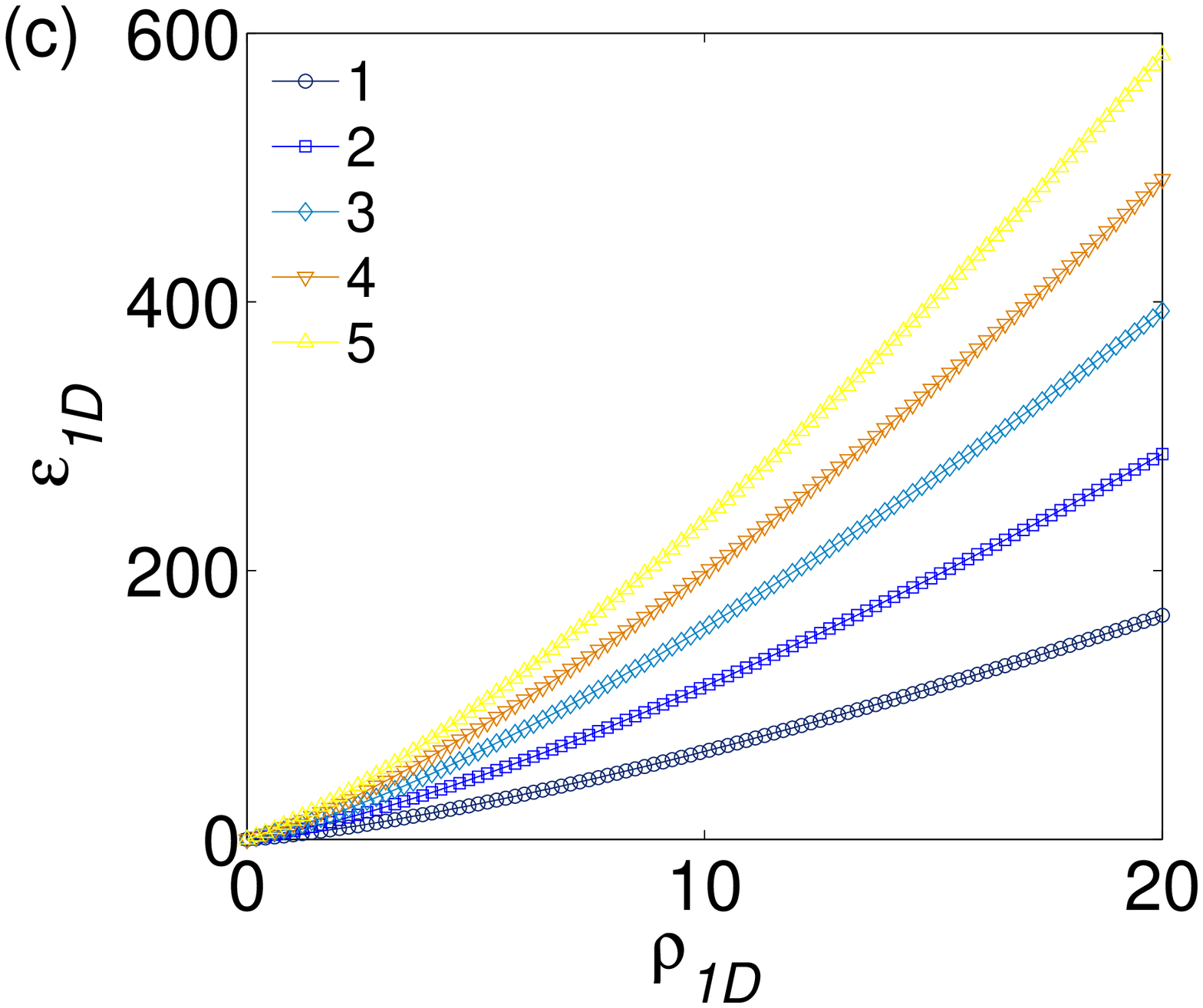}
\includegraphics{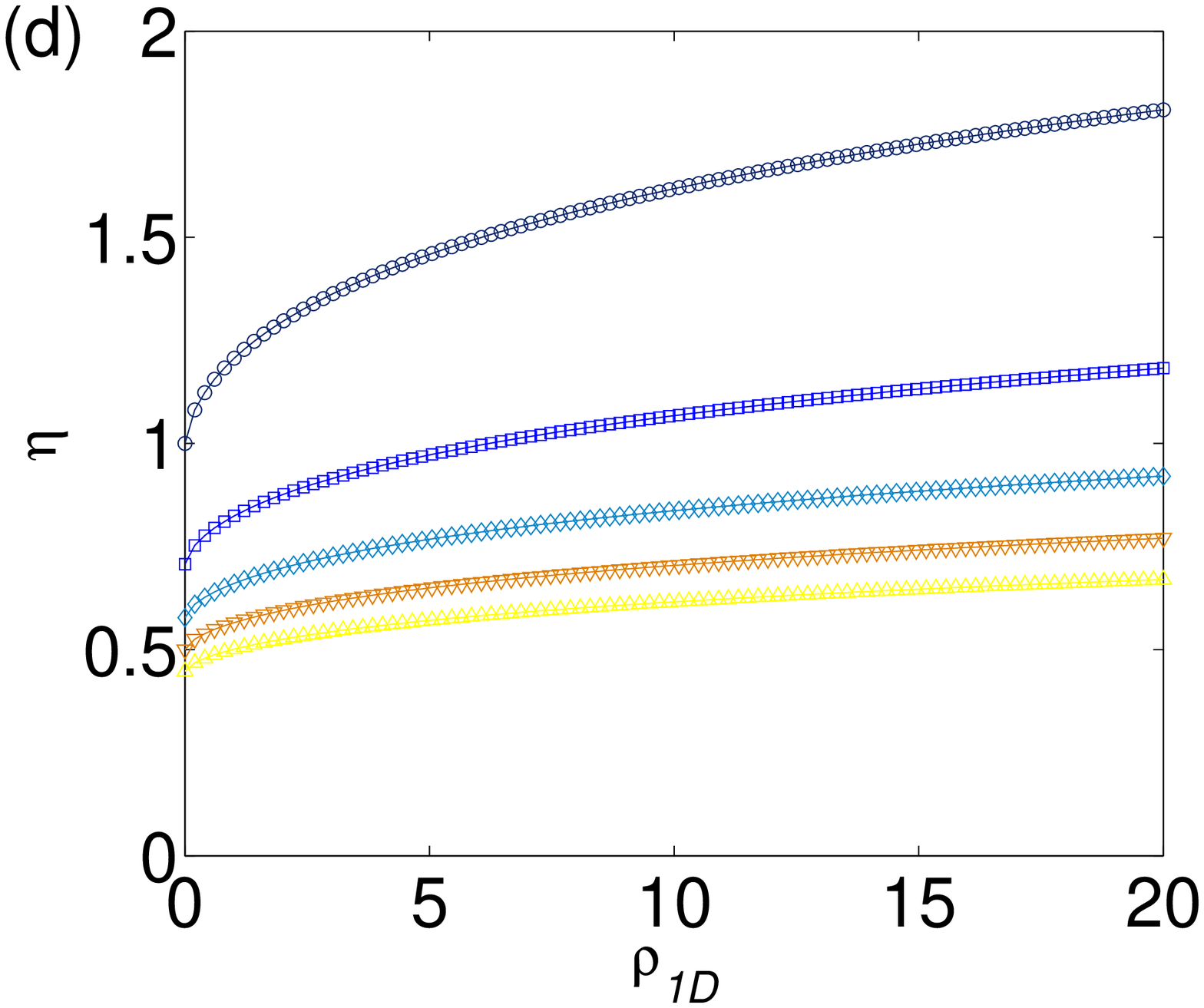}
}  }
\caption{(a) ${\protect\varepsilon }{_{\mathrm{1D}}}$ vs. ${{\protect\rho }_{1D}%
}$ and (b) ${\protect\eta }$ VS. ${{\protect\rho }_{\mathrm{1D}}}$. In both
plots the five curves correspond to ${a_{s}}/a_{c}=-0.2,-0.1,0,0.1$ and $0.2$%
, the confinement parameter being $\protect\varpi _{\mathrm{t}}=5$. (c) ${%
\protect\varepsilon }{_{\mathrm{1D}}}$ vs. ${{\protect\rho
}_{\mathrm{1D}}}$ and (d) ${\protect\eta }$ vs. ${{\protect\rho
}_{\mathrm{1D}}}$. In both plots the five curves correspond to
${\protect\varpi _{\mathrm{t}}}=1,2,3,4$ and $5$, and the scattering
length is fixed as ${a_{s}}/a_{c}=-0.1$.} \label{fig:7}
\end{figure}

The plots in Fig.~\ref{fig:7} show the dependence of the energy density , $%
\varepsilon _{\mathrm{1D}}$, and transverse width, $\eta $, on the 1D density, $%
\rho _{\mathrm{1D}}$, as obtained from Eqs.~(\ref{E-24}) and (\ref{E-26}),
respectively. The dependences are similar to their 2D counterparts shown in
Fig.~\ref{fig:1}, i.e., the energy density increases with the atomic
density, the increase being stronger at higher values of $\varpi _{\mathrm{t}%
}$, and with the strength of the repulsive interactions ($a_{s}>0$). The
similarity is also true for the dependence of the transverse width, $\eta $,
if compared to that for width, $\zeta $, in the 2D model. However,
estimating the 3D density, $n_{\mathrm{3D}}$, in the case of $\rho _{\mathrm{%
1D}}\approx \rho _{\mathrm{2D}}$ (which also implies that $\zeta \approx
\eta $), we find that the 3D density, as produced by the 1D model, is larger
by a factor $\sim 1/a_{c}$. Therefore, the range of values of $n_{\mathrm{3D}%
}$ corresponding to Fig.~\ref{fig:7} is much broader than in Fig.~\ref{fig:1}%
, implying that, using the strong 2D confinement, the unitary regime can be
achieved at much lower densities than with the 1D confinement, which is
explained by the fact that the Pauli-repulsion term is more important in the
latter case.

Again, we assume that the gas is subject to the action of the combination of
the harmonic trap and OL in the axial direction,

\begin{equation}
{U_{\mathrm{1D}}}=\varpi _{z}^{2}{\left( {Z-{Z_{0}}}\right) ^{2}}+{A_{z}}{%
\sin ^{2}}\left( {{\kappa _{z}}Z}\right) ,  \label{E-28}
\end{equation}%
where $\kappa _{z}\equiv 2\pi /\lambda _{z}$ is the OL wavenumber, and $%
\lambda _{z}/2$ the OL\ period. The results obtained for such axial
potentials are reported below.

\subsection{The trapped quasi-1D gas in the absence of the optical lattice}

\begin{figure}[tbp]
\centering
{\normalsize
\resizebox{0.8\textwidth}{!}{\includegraphics{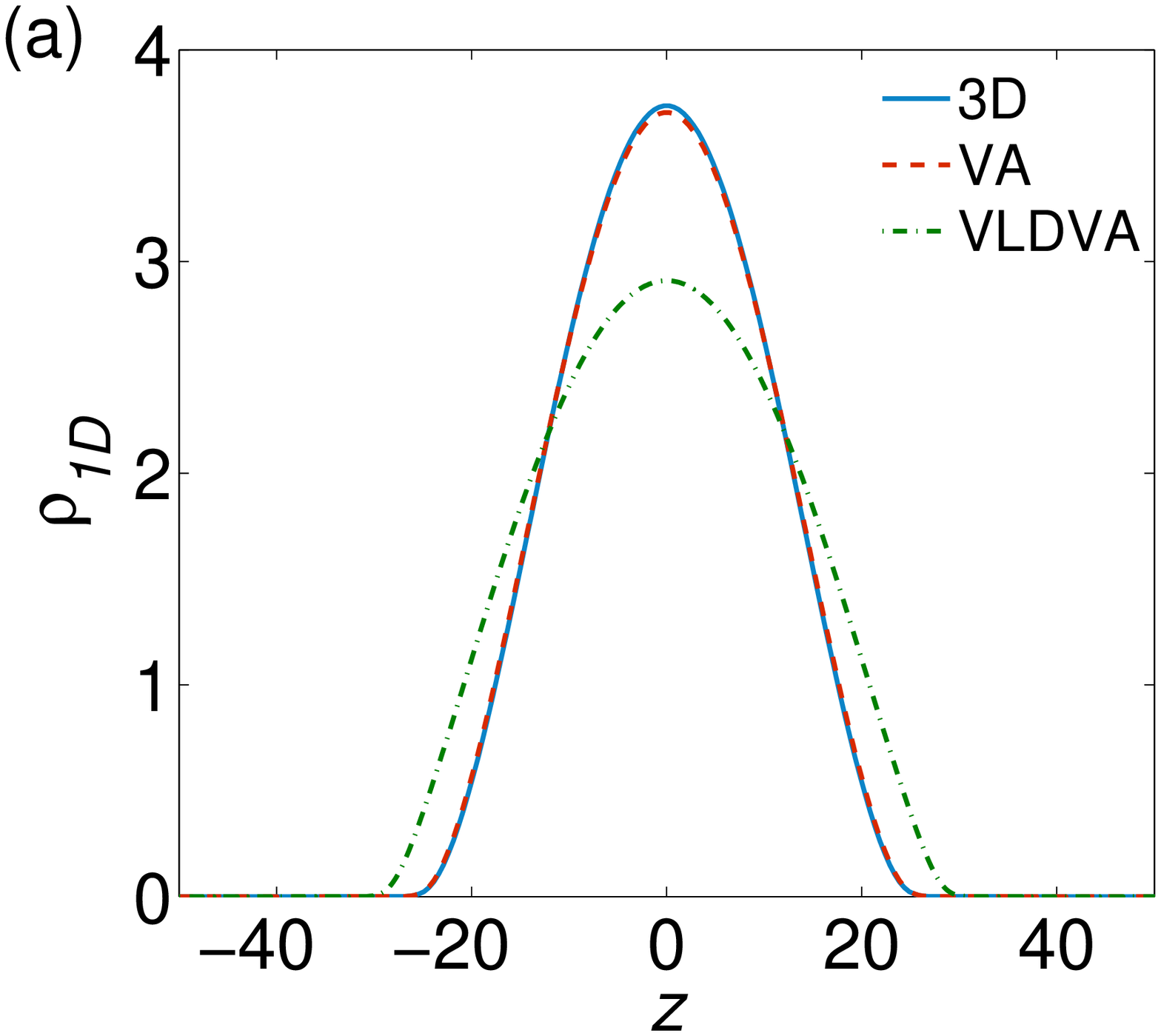}
\includegraphics{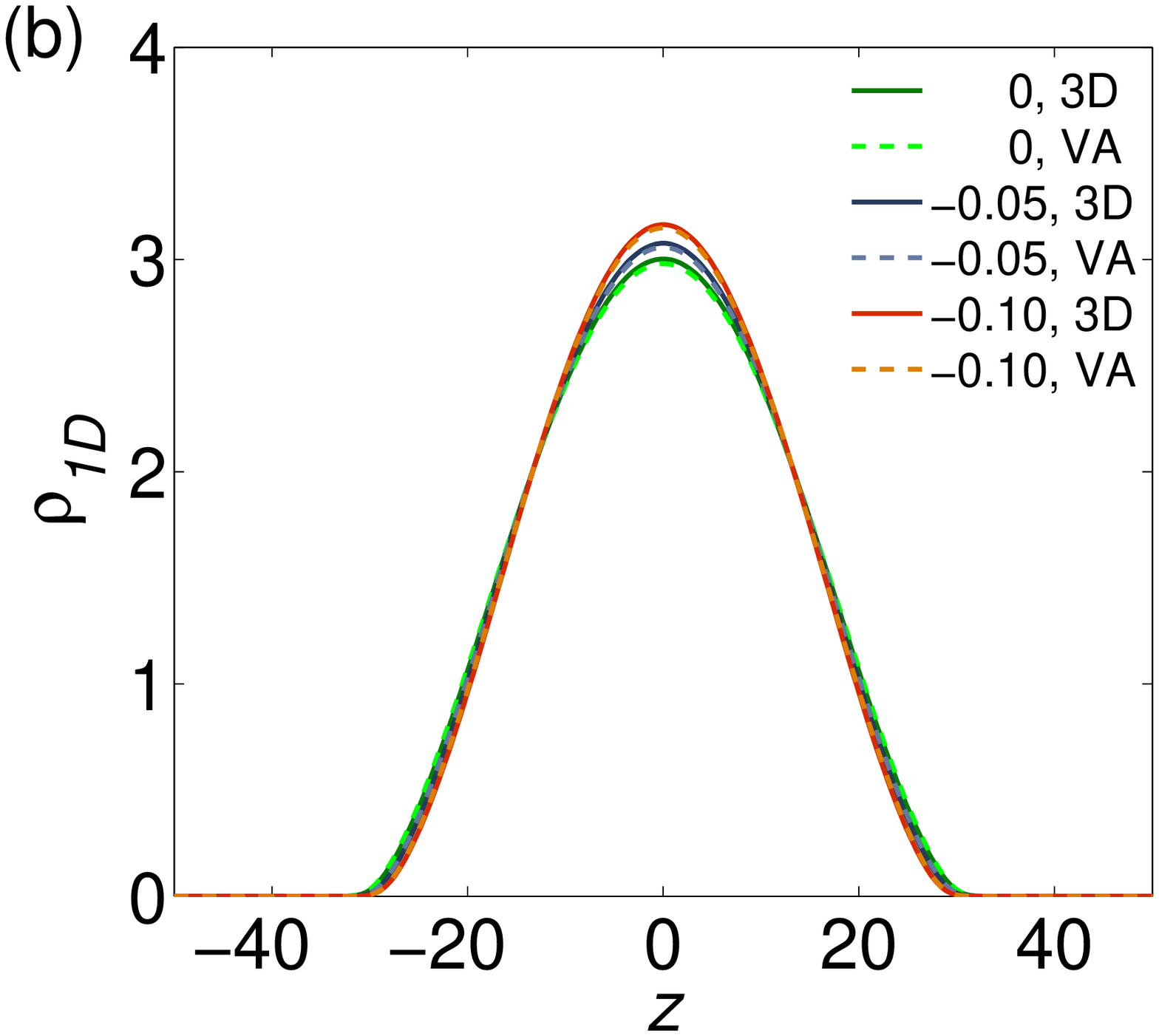}
}
\resizebox{0.8\textwidth}{!}{\includegraphics{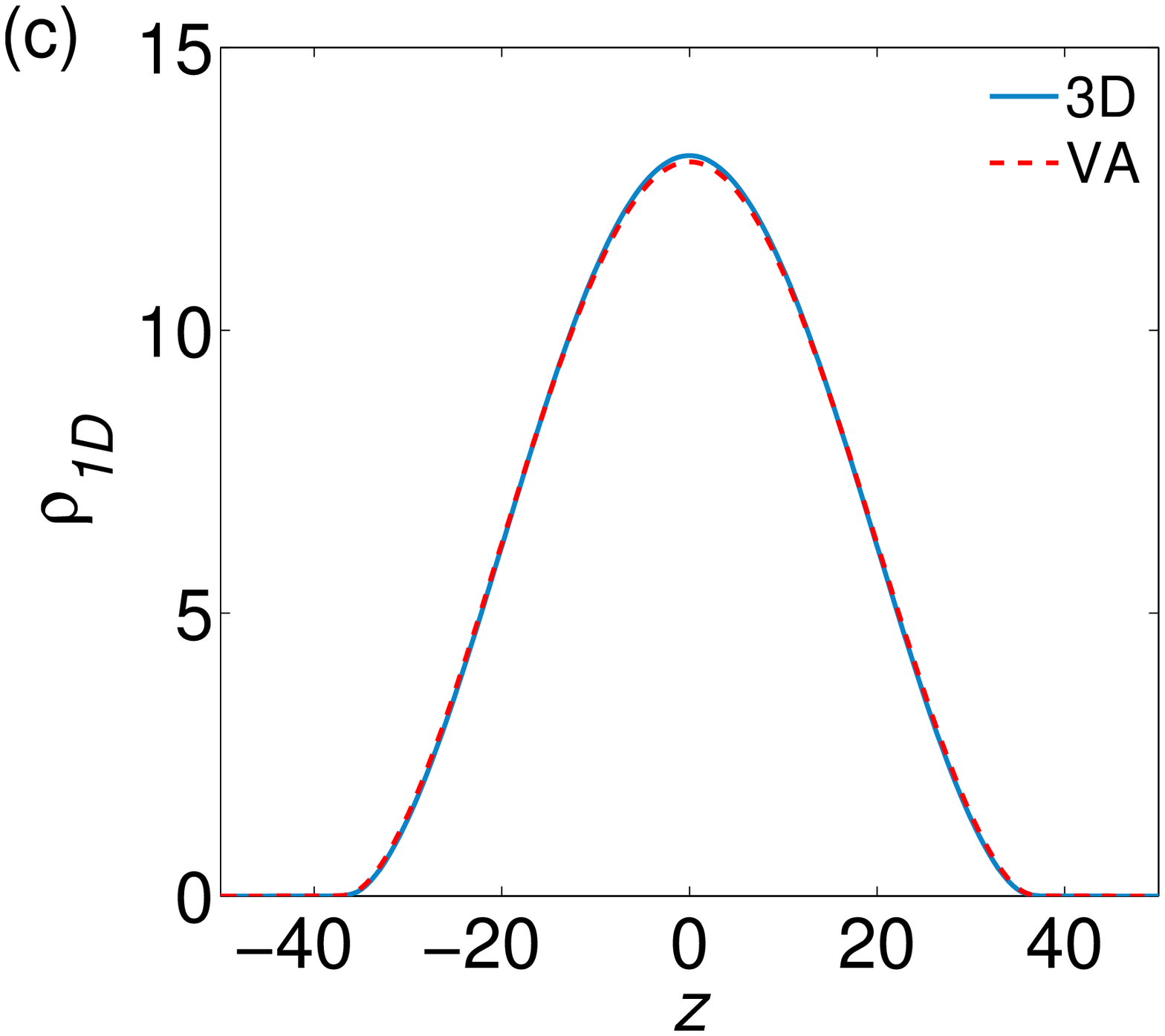}
\includegraphics{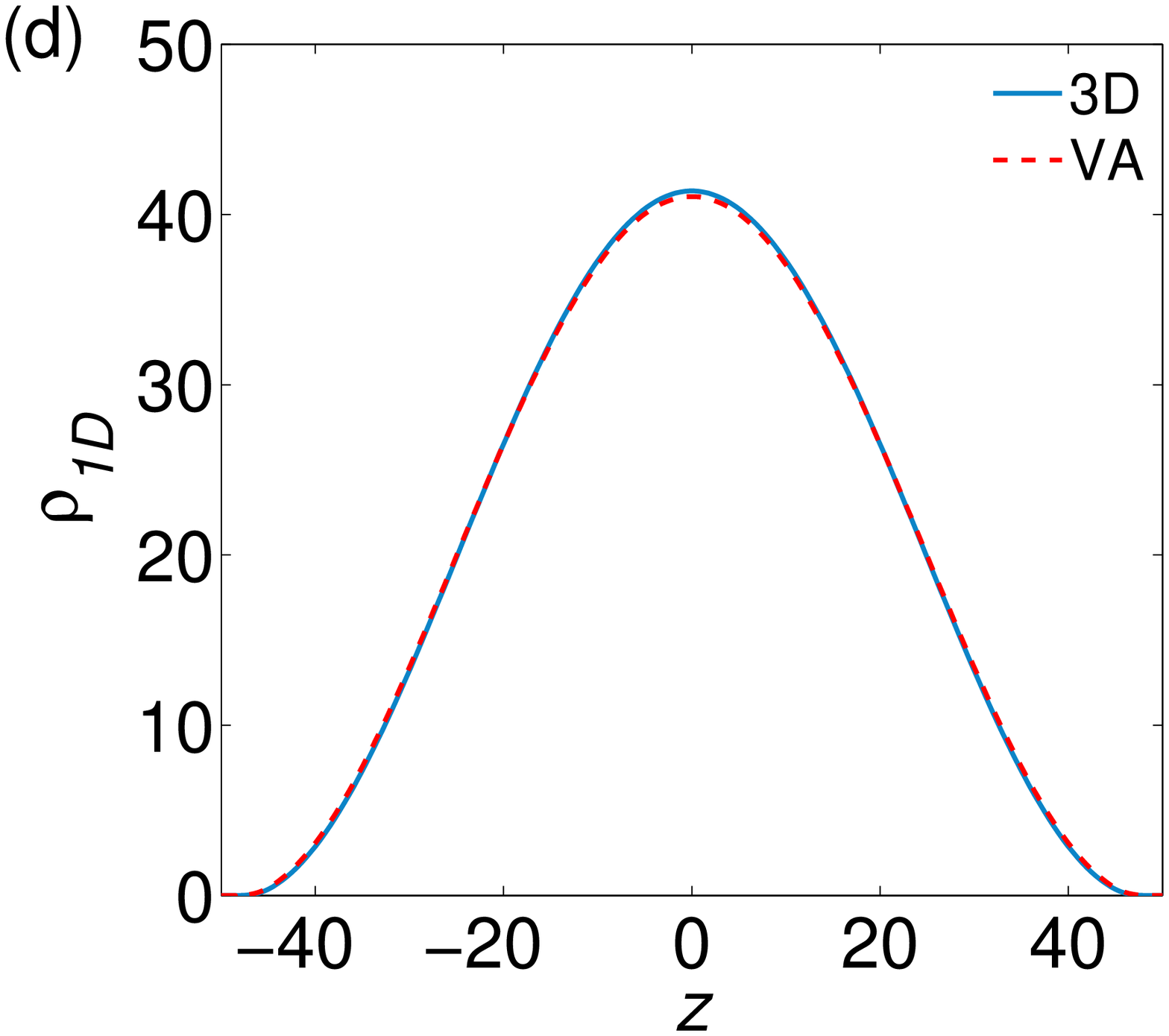}
}  }
\caption{(Color online) The 1D density profile, $\protect\rho _{\mathrm{1D}}$%
, as calculated through the full 3D solution (``3D"), and by dint of
the full 1D reduction based on the variational approximate (``VA")
(panel (a) also includes the limit case of the latter, corresponding
to the``very-low-density variational
approximation" (``VLDVA")). (a) Parameters are $N=100$, $%
a_{s}/a_{c}=0$, $\protect\varpi _{\mathrm{t}}=1$ and $\protect\varpi _{z}=0.1
$. (b) For three values of the scattering length, $a_{s}/a_{c}=0,-0.05$ and $%
-0.1$, where the other parameters are $N=100$, $\protect\varpi _{\mathrm{t}%
}=2$ and $\protect\varpi _{z}=0.1$. (c) $N=500$, and (d) $N=2000$, for the
other parameters $a_{s}/a_{c}=-0.05$, $\protect\varpi _{\mathrm{t}}=2$ and $%
\protect\varpi _{z}=0.1$.}
\label{fig:8}
\end{figure}

As in the previous section, it is first necessary to check the accuracy of
the approximation. To this end, in Fig.~\ref{fig:8}(a) we compare the 1D
density, $\rho _{\mathrm{1D}}$, as produced by three different methods: the
full 3D equation (\ref{E-2}), the full 1D approximation, and its low-density
limit, see Eq. (\ref{E-27}). Similar to the 2D case, we observe an excellent
agreement between the 3D equation and the full 1D approximation, both
significantly disagreeing with the low-density limit. In particular, Fig.~%
\ref{fig:8}(b) demonstrates that the approximation remains accurate in the
presence of the interactions too. On the other hand, Figs.~\ref{fig:8}(c)
and (d) show that an increase in the number of atoms does not affect the
agreement between the 1D and 3D simulations.

\begin{figure}[tbp]
\centering
{\normalsize
\resizebox{0.8\textwidth}{!}{\includegraphics{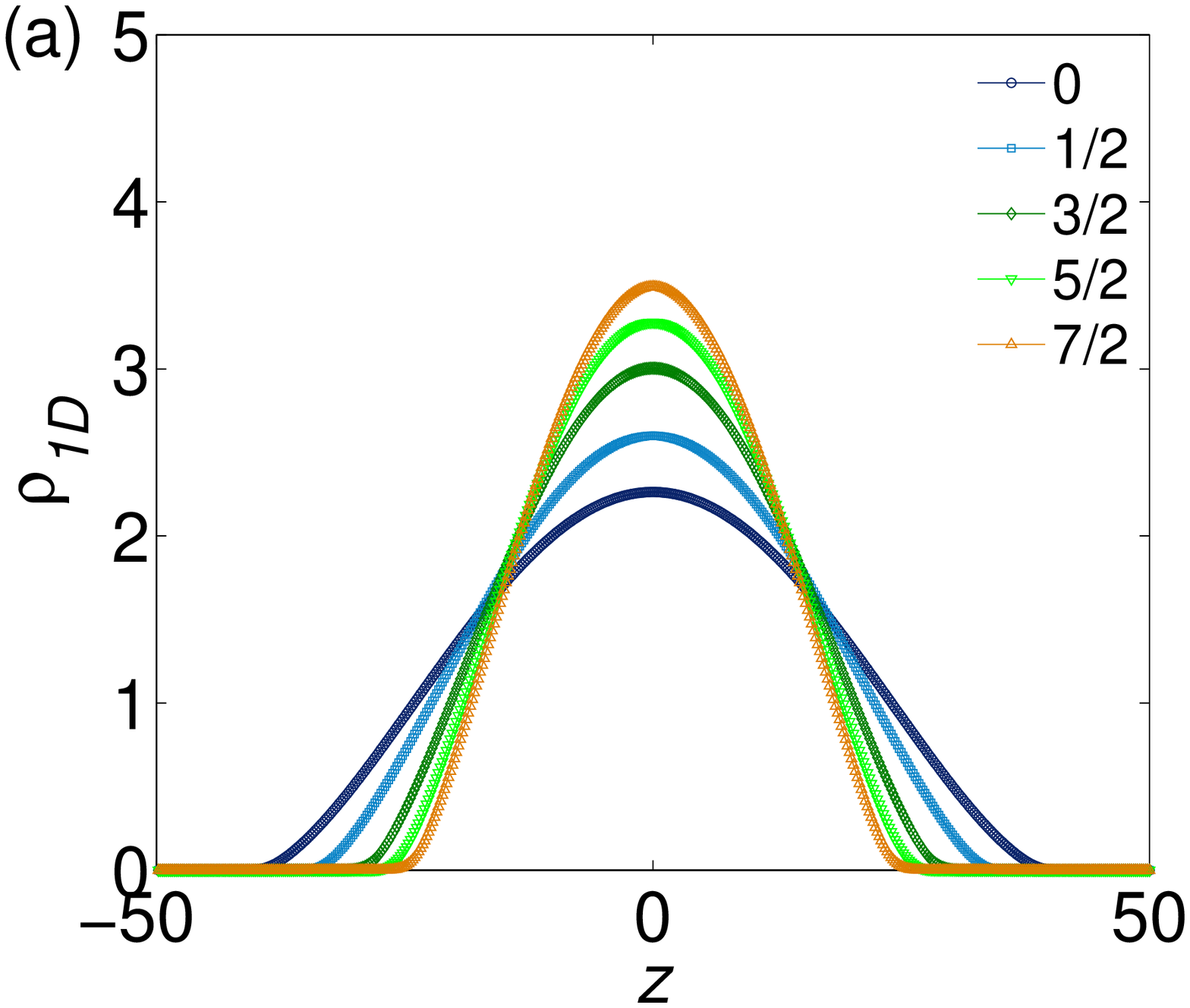}
\includegraphics{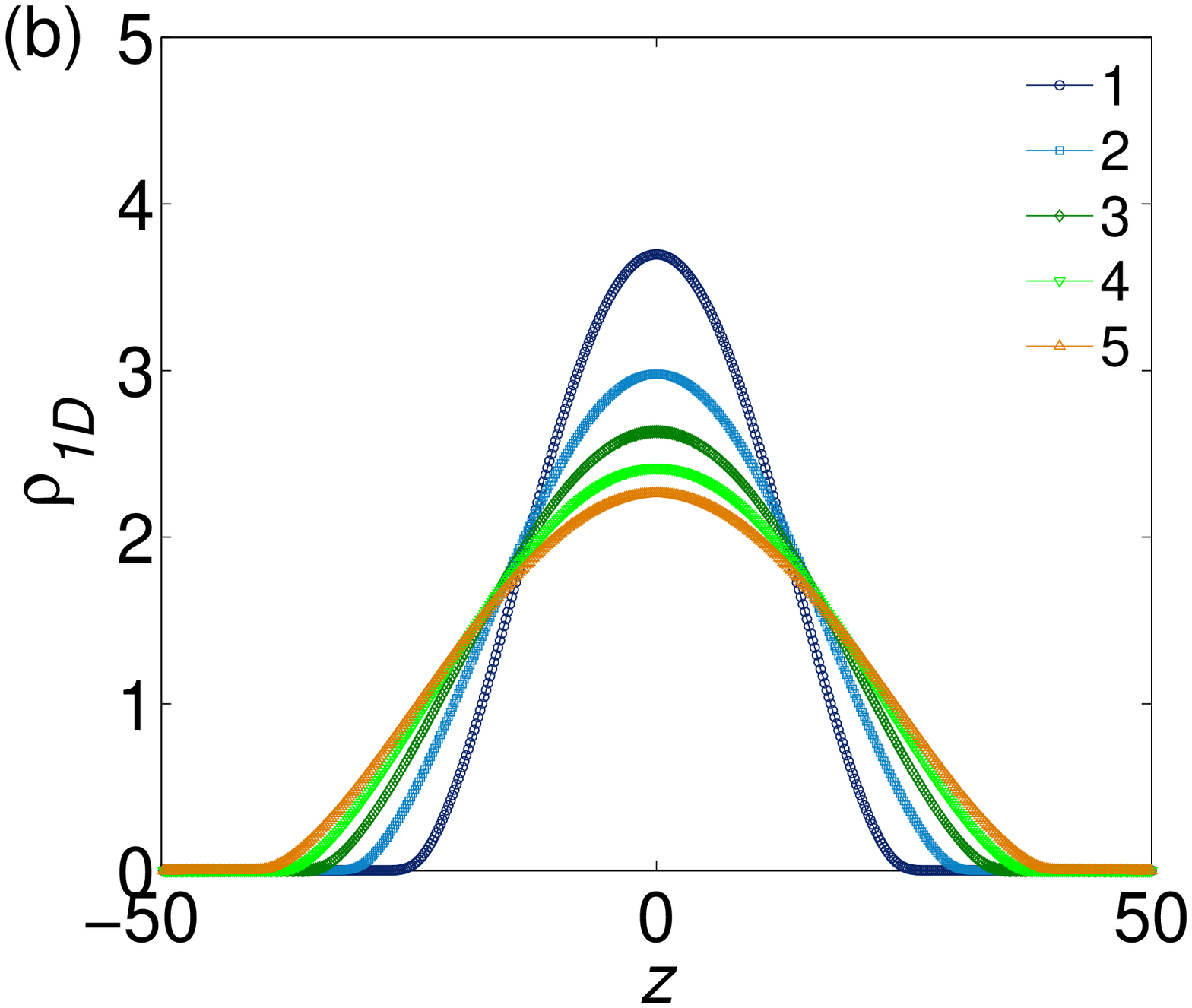}
}  }
\caption{(Color online) (a) The 1D density profile. Five 1D density profiles
correspond to $s=0,1/2,3/2,5/2$, $7/2$ (as indicated in the inset) and $%
\protect\varpi _{\mathrm{t}}=5$. (b) The same, with the five curves
corresponding to $\protect\varpi _{\mathrm{t}}=1,2,3,4$, $5$ (as per the
inset) and $s=0$. The parameters are $N=100$, $a_{s}/a_{c}=0$ and $\protect%
\varpi _{z}=0.1$.}
\label{fig:9}
\end{figure}

Figure~\ref{fig:9}(a) shows profiles of the 1D density, $\rho _{\mathrm{1D}}$%
, for five different values of spin $s$, in the presence of the confining
harmonic-oscillator potential acting in the axial direction ($z$). The
smallest width of the profile is observed at largest $s$, which is a result
of the relatively weak Pauli repulsion versus the attractive interactions,
as a consequence of the lower spin degeneracy. Figure~\ref{fig:9}(b) shows
axial expansion of the trapped Fermi gas with the increase of the strength
of the transverse confinement ($\varpi _{\mathrm{t}}$). This trend is a
consequence of the rapid enhancement of the Pauli repulsion.

\subsection{Effects of the axial optical lattice}

\begin{figure}[tbp]
\centering
{\normalsize
\resizebox{0.8\textwidth}{!}{\includegraphics{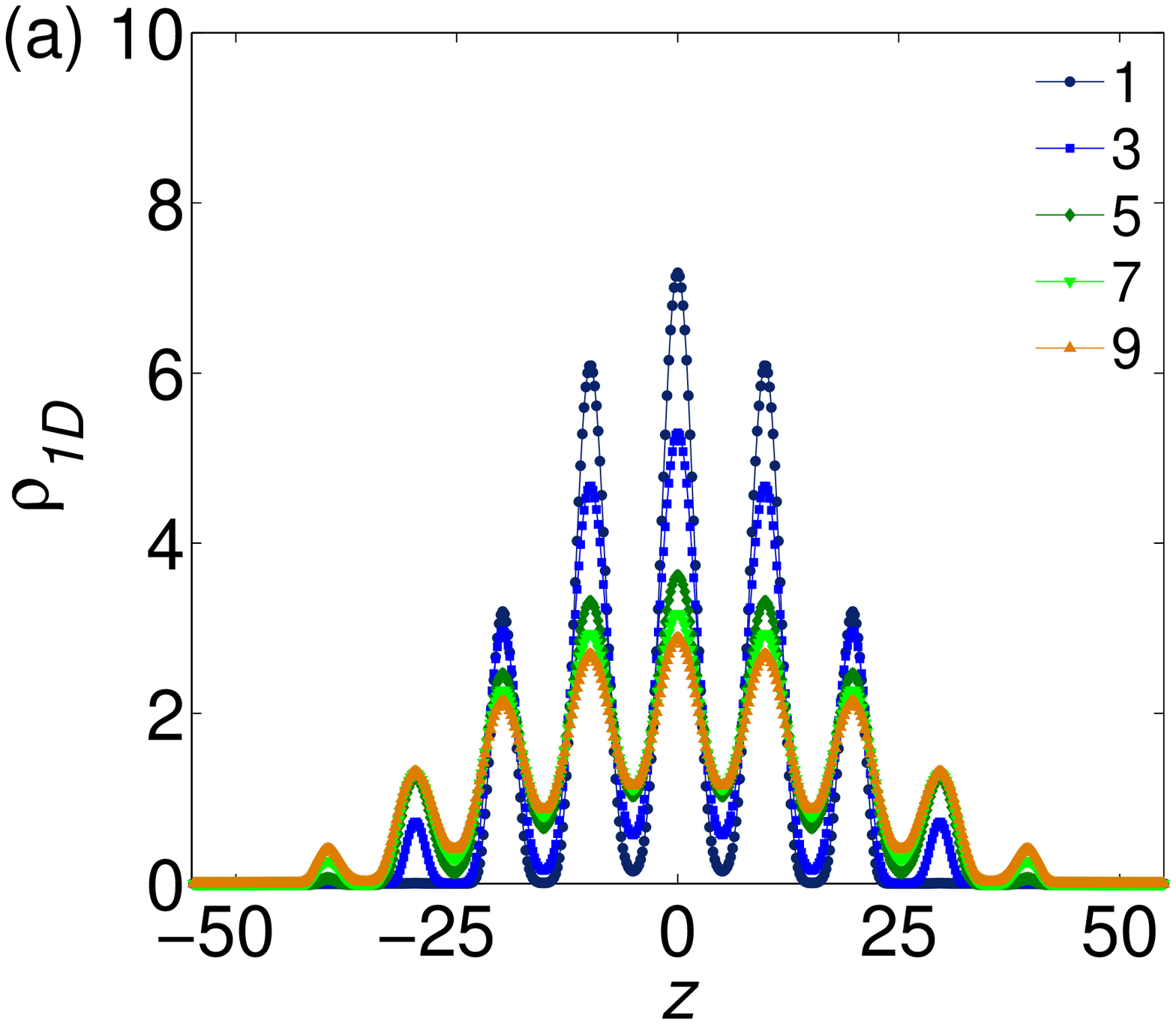}
\includegraphics{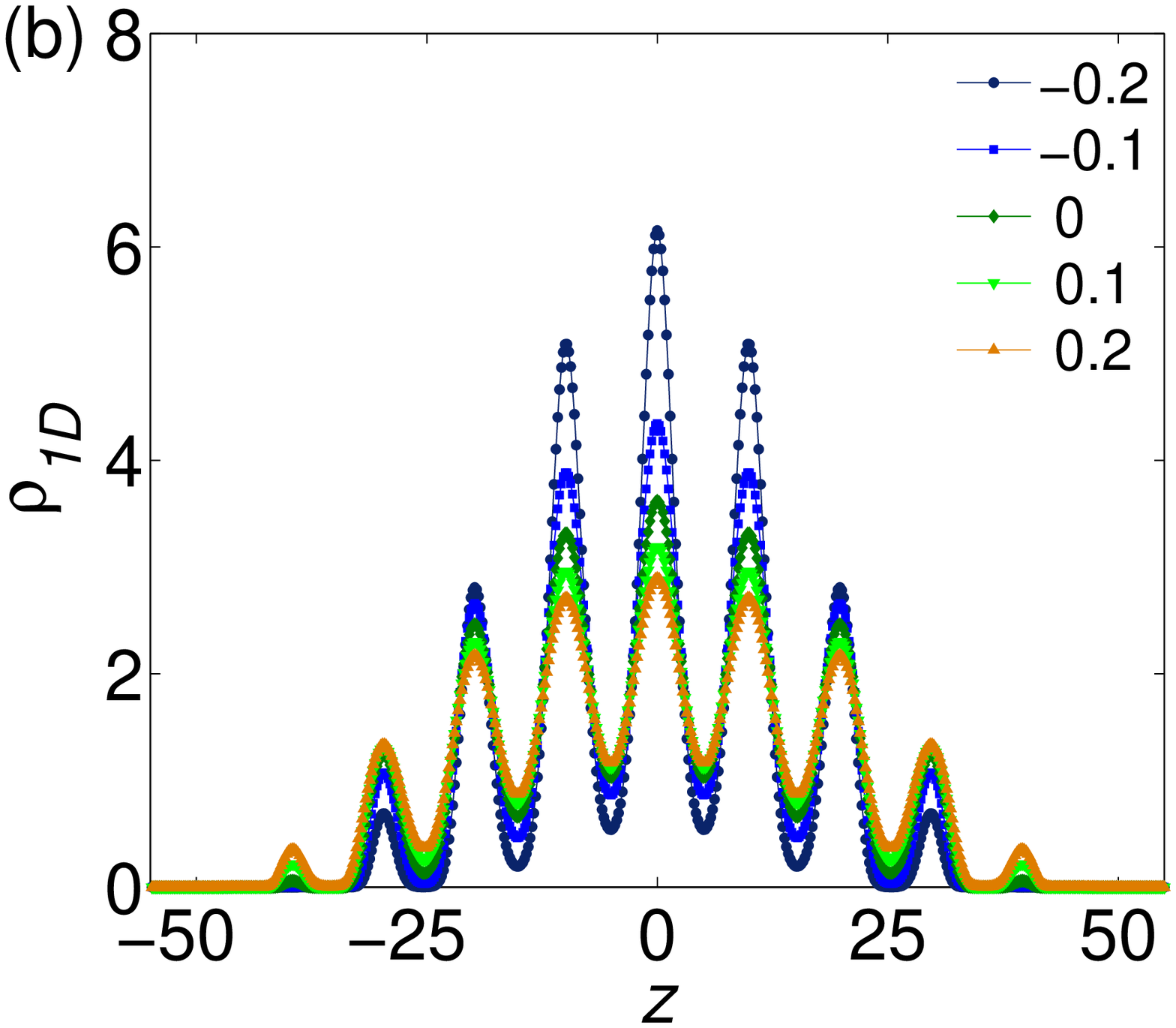}}  }
\caption{(Color online) One-dimensional density profiles of the Fermi gas
trapped in the OL with amplitude (depth) $A_{z}=10$ and double period $%
\protect\lambda _{z}=20$. (a) The five curves correspond to the confinement
strength $\protect\varpi _{\mathrm{t}}=1,3,5,7,9$ (as indicated in the top
right inset). The parameters are $N=100$, $G=0$, and $\protect\varpi _{z}=0.1
$. (b) The five curves correspond to scattering lengths $%
a_{s}/a_{c}=0.2,0.1,0,-0.1,-02$, as indicated in the panel. The parameters
are $N=100$, $\protect\varpi _{\mathrm{t}}=5$, and $\protect\varpi _{z}=0.1$.
}
\label{fig:10}
\end{figure}

Typical 1D-density patterns for the gas trapped in the OL are displayed in
Fig.~\ref{fig:10}(a). Note that the increase of the confinement strength
leads to attenuation of the central peak and generation of new side peaks,
which is consistent with the elongation of the gas observed in Fig.~\ref%
{fig:9}(b). The lowering of the central peak is accompanied by the increase
of the density between peaks, resulting in the reduction of their
visibility. A similar behavior is observed in Fig.~\ref{fig:10}(b), where
the repulsive inter-atomic interaction makes the trapped state elongated
along $z$, resembling the elongation induced by the strengthening transverse
confinement, and vice versa for the attractive interaction. Figure~\ref%
{fig:11} shows the strong dependence of the 1D density profile $\rho _{%
\mathrm{1D}}$ on the OL's (double) period $\lambda _{z}$, for a fixed
strength of the confinement potential. It is clear that the number of peaks
in the 1D pattern is larger for smaller $\lambda _{z}$. The increase of $%
\lambda _{z}$ entails disappearance of peripheral peaks at particular values
of $\lambda _{z}$. The present 1D description makes it possible to identify
these critical values.

\begin{figure}[tbp]
\centering
{\normalsize
\resizebox{0.5\textwidth}{!}{\includegraphics{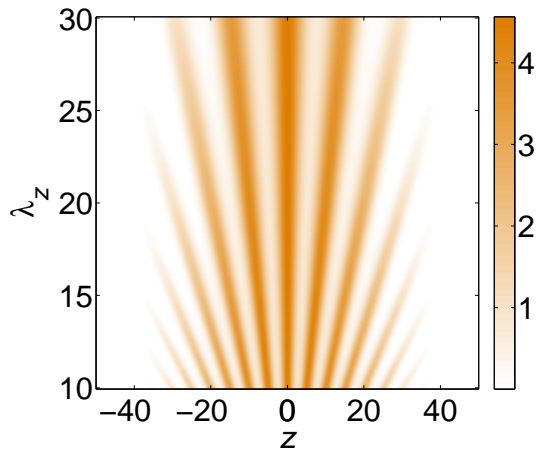}
}  }
\caption{(Color online) One-dimensional profiles of the Fermi gas trapped in
the axial optical lattice with amplitude $A_{z}=10$. The plot shows density $%
\protect\rho _{\mathrm{1D}}$ versus the double lattice's period $\protect%
\lambda _{z}$. The parameters are $N=100$, $G=0$, $\protect\varpi _{\mathrm{t%
}}=5$, $\protect\varpi _{z}=0.1$.}
\label{fig:11}
\end{figure}

\begin{figure}[tbp]
\centering
{\normalsize
\resizebox{0.8\textwidth}{!}{\includegraphics{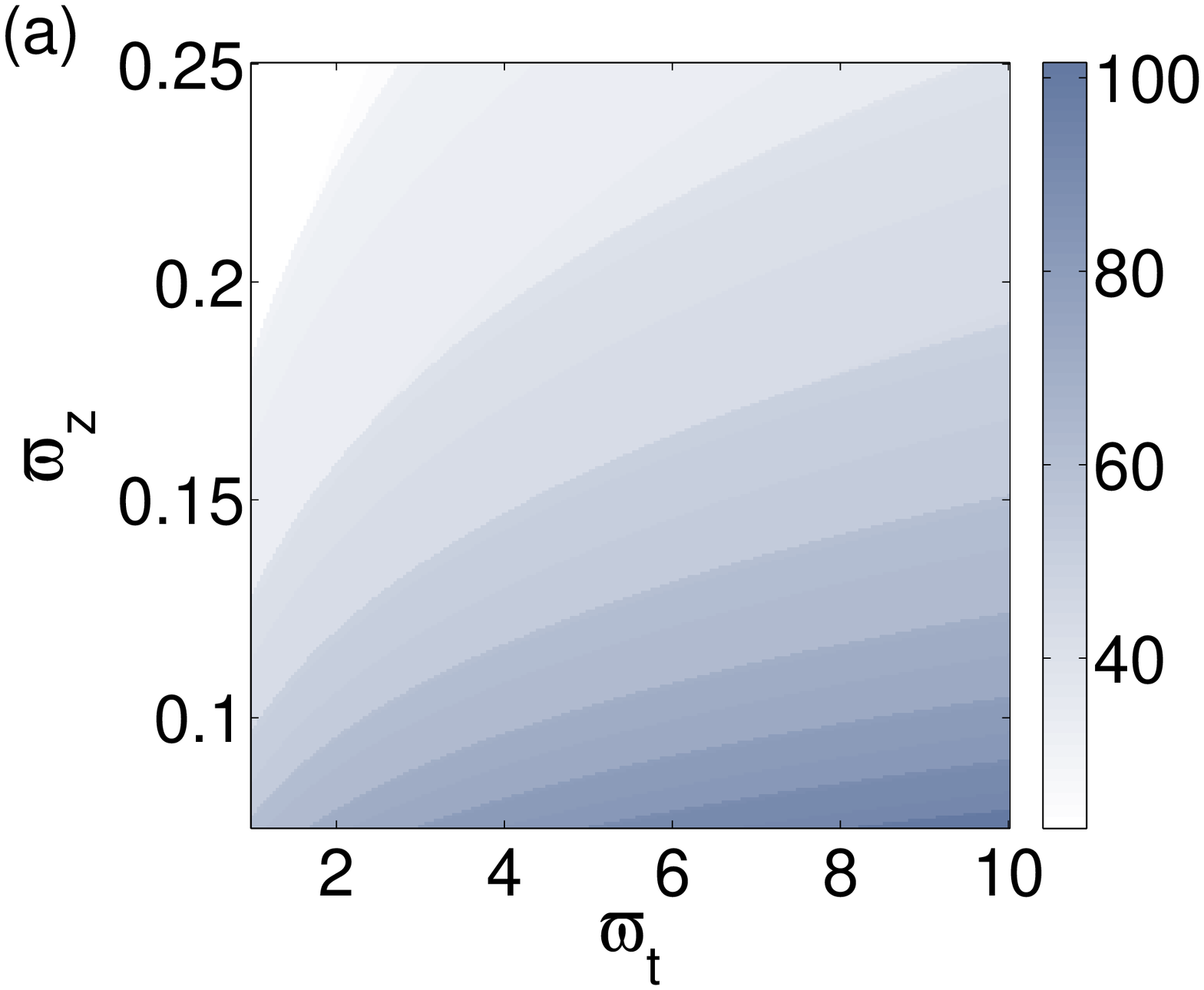}
\includegraphics{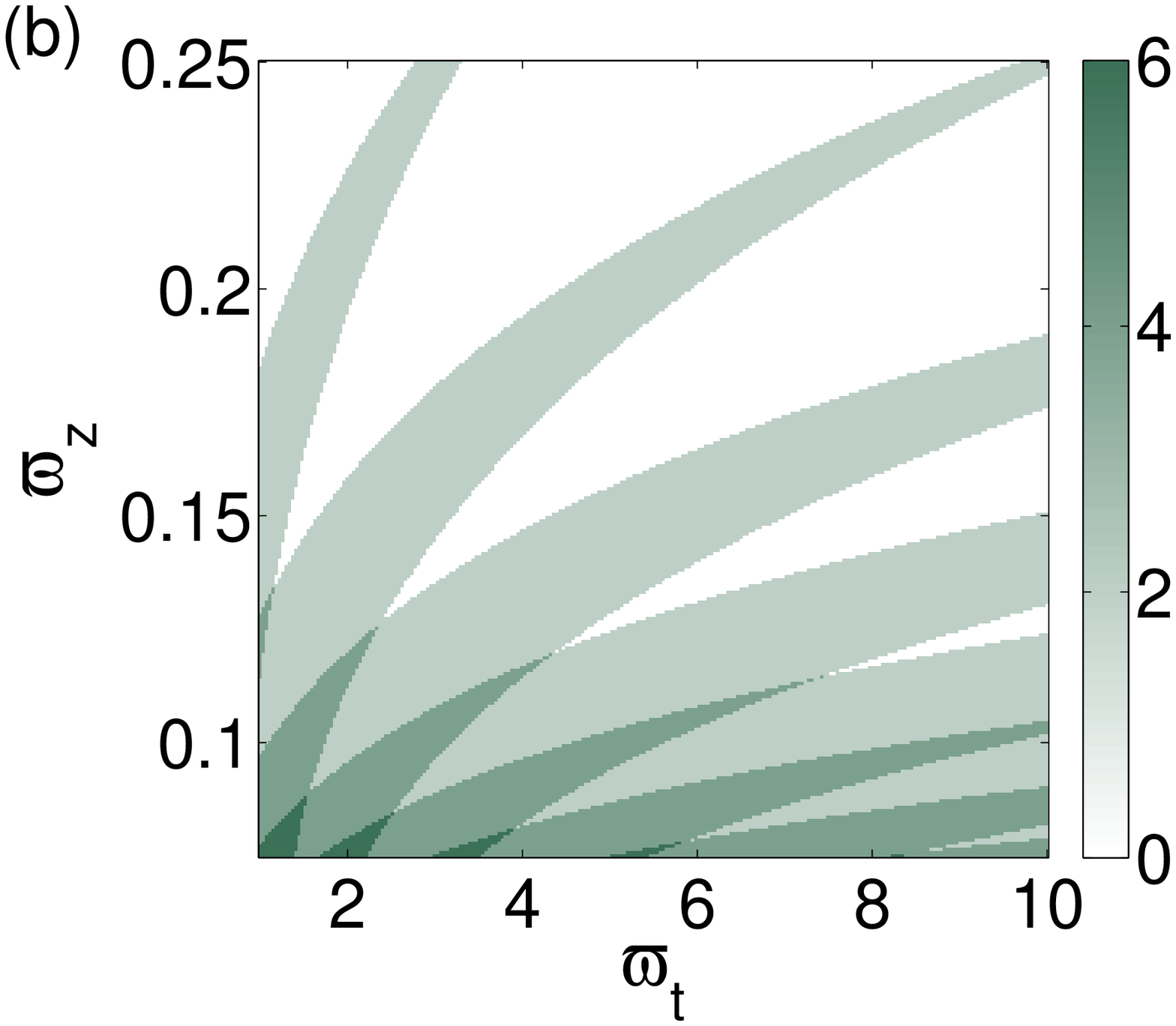}
}  }
\caption{(Color online) Diagrams in the plane of the trapping frequencies, $%
\left( \protect\varpi _{\mathrm{t}},\protect\varpi _{z}\right) $: (a) width $%
\Delta _{z}$ of the area wherein $\protect\rho _{\mathrm{1D}}>0.05$; (b) the
number of density peaks at the periphery of the trapped state, which are
disconnected from the central peak. The parameters are $N=100$, $%
a_{s}/a_{z}=-0.05$, $A_{z}=10$, $\protect\lambda _{z}=10$.}
\label{fig:12}
\end{figure}

To further analyze the ground state of the Fermi gas in presence of the OL,
we define the width of the trapped state, $\Delta _{z}$, as the size of the
area with density $\rho _{\mathrm{1D}}>0.05$. Figure~\ref{fig:12} presents
diagrams for characteristics of the ground state in the plane of the
transverse and axial trapping frequencies, $\left( \varpi _{\mathrm{t}%
},\varpi _{z}\right) $. The jumps observed in Fig.~\ref{fig:12}(a)
correspond to the emergence or disappearance of pairs of density peaks.
Since the wavelength is $\lambda _{z}=10$, the emergence or disappearance of
the peak pair gives rise, respectively, to the increase or decrease of the
width by $\Delta _{z}=10$. Thus, the observed minimum width ($\Delta _{z}=20$%
) corresponds to the profile with five peaks, and the maximum width ($\Delta
_{z}=110$) is represented by the profile with twenty three peaks. Note that,
as seen in Fig.~\ref{fig:10}, the density between peaks may be very small.

Further, we define that two peaks as being ``connected" if
the minimum density between them is greater than $0.05$. Accordingly, Fig.~%
\ref{fig:12}(b) shows the number of peaks ``disconnected" from the
central core. This characteristic is found by subtracting the number
of the connecting links between the peaks, defined as said above,
from the total number of the observed peaks (excluding the central
one). The diagram features four regions covered by similar colors,
which correspond to four possible values of the number of isolated
peaks, $0,2,4$ and $6$. These
regions alternate with the variation of both $\varpi _{\mathrm{t}}$ and $%
\varpi _{z}$, demonstrating that the number of the isolated peaks may be
maintained while varying the total number of the observed peaks.

\section{Conclusions}

\label{sec:4} We have demonstrated that the 1D and 2D reductions of the 3D
mean-field description of the degenerate Fermi gas, based on the Gaussian
variational ansatz, provide very accurate results for the ground states of
the gas in the disk- and cigar-shaped traps. The so derived reduced
equations are useful, in particular, for modeling the BCS regime at low
atomic density and large spatiotemporal variations of the external
potential. For the 1D case, the reduced equations provide results requiring
the use of modest computational resources and allow one to quickly explore a
vast volume of the parameter space. In the 2D case, the necessary simulation
time is still significantly lower than what is necessary for the 3D
simulations. For the spatially uniform ground states in both dimensions, we
have studied the width of the Fermi gas in the confinement dimensions, and
the energy density as a function of the atomic density. In the 2D case, we
have found that the energy density attains a maximum at a particular value
of the atomic density, and the maximum shifts to lower values as the
attractive interactions get stronger. While the analysis is approximate due
to limitations of the model, it shows the correct dependence on the
scattering length, confinement strength, and the phase-space degeneracy
(i.e., the fermionic spin, $s$).

For the gas trapped in the 2D OL, the analysis predicts the number of
visible peaks in the density profile as a function of the scattering length
and confinement strength. It was concluded that small variations of the
latter parameter produce significant changes in the size and distribution of
the peaks. In the 1D setting including the axial OL, we have produced the
phase diagrams for the Fermi gas, varying the confinement parameters.
Discrete leaps in the diagrams correspond to the appearance and
disappearance of a pair of peripheral peaks in the density profile. In fact,
the number of peaks does not change in a wide range of the confinement
parameters.

In general, our model is able to include any type of the confinement
and potential lattices in the 1D and 2D settings, making the
description of the ground state quite simple. In particular, we
considered the example of the hexagonal superlattice built as a
superposition of two triangular lattices. The latter setting may be
essential for modeling configurations relevant to solid-state
physics. The same approach may be further applied to superlattices
built of hexagonal OLs, with the objective to emulate graphene-like
superlattice. Beyond the study of the ground states, it may be
interesting to build quasi-2D modes with embedded vorticity, cf.
Ref. \cite{Malomed2009}. On the other hand, the reduced equations can be used to
study the dynamics of the Fermi gas under the action of external
potentials slowly varying in time, similar to how it was done in BEC
models. Finally, the same type of the description, including, if
necessary, equations for the Bose gas, may be used to predict the
ground state and analyze the low-energy dynamics of Fermi-Fermi and
Bose-Fermi mixtures.

\section*{Acknowledgements}

We thank Harald Pleiner for his critical reading of the manuscript. D.L.
acknowledges partial financial support from FONDECYT 11080229, 1120764,
Millennium Scientific Initiative, P10-061-F, Basal Program Center for
Development of Nanoscience and Nanotechnology (CEDENNA), Performance
Agreement Project UTA/ Mineduc and UTA-project 8750-12. P.D. acknowledges
the CONICYT PhD program fellowship. I.S. acknowledges partial financial
support from FONDECYT 1100287, Conicyt-DFG grant 084-2009. B.A.M.
appreciates partial support from the German-Israel Foundation (grant No.
I-1024-2.7/2009) and Binational (US-Israel) Science Foundation (grant No.
2010239).

\end{document}